\documentclass[a4paper,10pt]{article}

\usepackage{graphicx}
\usepackage{amssymb}
\usepackage{amsmath}
\usepackage[T1]{fontenc}
\usepackage{color}

\usepackage[varg]{txfonts}	% fonts for integrals etc

\definecolor{oneblue}{rgb}{0,0,0.65}
\definecolor{onered}{rgb}{0.15,.65,0.75}

\def\url#1{\textcolor{blue}{\underline{#1}}}	% web references
\def\corr#1{#1}    		% no corrections

\def\b#1{\textcolor{oneblue}{#1}} 
\newcommand{\degK}{${\,}^\circ K{\,}$}          % Temperature (Kelvin)

 \large\normalsize	 %double
\begin{document}

%\renewcommand{\baselinestretch}{1.1} \small\large

% ------------------------------------------------------------
% Title page
% ------------------------------------------------------------

%\title{Neuronal magnetic fields in complex extracellular media}
%\title{Generalized magnetic cable formalism for neurons}

\title{Generalized cable formalism to calculate the magnetic
    field of single neurons and neuronal populations}

% Short title (if needed)
% \subtitle{Generalized magnetic cable equations}

\author{Claude Bedard and Alain Destexhe}

\date{{\it Physical Review E}, in press, \today}

\maketitle

\section*{Abstract}

Neurons generate magnetic fields which can be recorded with
macroscopic techniques such as magnetoencephalography.  The theory
that accounts for the genesis of neuronal magnetic fields involves
dendritic cable structures in homogeneous resistive extracellular
media.  Here, we generalize this model by considering dendritic cables
in extracellular media with arbitrarily complex electric properties.
This method is based on a multi-scale mean-field theory where the
neuron is considered in interaction with a ``mean'' extracellular
medium (characterized by a specific impedance).  We first show that,
as expected, the generalized cable equation and the standard cable
generate magnetic fields that mostly depend on the axial current in
the cable, with a moderate contribution of extracellular currents.
Less expected, we also show that the nature of the extracellular and
intracellular media influence the axial current, and thus also
influence neuronal magnetic fields.  We illustrate these properties by
numerical simulations and suggest experiments to test these findings.

\section{Introduction}

Neuronal magnetic activity is usually measured through
magnetoencephalogram (MEG) signals, which are recorded by using
sensitive Superconducting Quantum Interference Device (SQUID)
detectors.  These sensors operate at very low temperatures
\corr{(4\degK)}, and must necessarily be located centimeters away from
the human scalp \cite{Weinstock1996}.  Because of the macroscopic
aspect of SQUID measurements, it is usually assumed that the
underlying sources are ``macroscopic dipoles'' produced by the
synchronized activity of thousand of neurons in a small region of
cortex \cite{Hamailainen1993}.

However, since a few years, many efforts were devoted to building
magnetic sensors of another kind, which are based on the Giant
Magneto-Resistance (GMR) effect in spin electronics
\cite{Freitas2012}.  Such sensors have the advantage of being able to
work at physiological temperatures, and they can be miniaturized, so
it is possible to build ``magnetrodes'' \cite{Pannetier2011}, the
magnetic equivalent of a micro-electrode.  Such devices are aimed to
record microscopically, the activity of a small group of neurons.
While the theory exists for macroscopic SQUID measurements and
macroscopic neuronal sources \cite{Hamailainen1993}, the theory to
explain the genesis of magnetic fields by single neurons has been very
scarsely developed \cite{Murakami}.  This is the first motivation of
the present study.

The second motivation follows from a controversy in the literature
about the role and properties of the extracellular medium around
neurons \cite{Buzsaki,DesBed2013}.  The ``standard'' model of the
genesis of the extracellular local field potential (LFP) assumes that
the neurons are dipolar sources embedded in a resistive (Ohmic)
extracellular medium.  While some measurements seem to confirm this
hypothesis~\cite{Logo}, other measurements revealed a marked frequency
dependence of the extracellular
resistivity~\cite{Gabriel1996a,Gabriel1996b}, which indicates that the
medium is non-resistive \corr{ or non-Ohmic\footnote{\corr{In a
      non-Ohmic medium, the differential Ohm's law
      (\b{$\vec{j}^f=\sigma\vec{E}=cst*\vec{E}$}) does not apply.}}}.
Indirect measurements of the extracellular impedance, as well as the
spectral analysis of LFP signals, also indicate deviations from
resistivity~\cite{BedDes2006a,Baz2011,BedDes2010,Deg2010}.  Such
deviations can be explained by phenomena like ionic
diffusion~\cite{BedDes2011a}, which reproduce the correct
frequency-scaling of LFP signals, In addition, there is also
evidence~\cite{Riera} that multipolar components are not sufficient to
explain the data, but that a strong monopolar component should be
taken into account.

These controversies have important consequences, because if the
extracellular medium is non-resistive, several fundamental theories of
neural dynamics, such as the well-known cable theory of
neurons~\cite{Rall1962,Rall1995} or the Current-source density
analysis~\cite{Mitzdorf}, are incorrect and need to be reformulated
accordingly~\cite{BedDes2011a,BedDes2013}.  The same considerations
may also hold for the genesis of the magnetic fields, as the current
theory~\cite{Hamailainen1993} also assumes that the medium is
resistive.

In the present paper, our aim is to build a neuron model to generate
electromagnetic fields based on first principles, and that does not
make any a priori assumption, such as the nature of the impedance of
the extracellular medium.  However, to this end, we cannot use the
classic cable formalism, which was initially developed by
Rall~\cite{Rall1962}.  Although this formalism has been one of the
most successful formalism of theoretical neuroscience, explaining a
large range of phenomena \cite{Rall1995,Tuckwell,Wu,Koch}, it is non
valid to describe neurons in non-resistive media.  To palliate to this
difficulty, we have recently generalized cable theory to make it valid
for neurons embedded in media with arbitrarily complex electrical
properties~\cite{BedDes2013}.  In the present framework, we will use
this generalized cable theory which will be extended to calculate
neuronal magnetic induction and electric potential in extracellular
space.

We start by outlining a generalized theoretical formalism to calculate
the magnetic field around neurons, and we next illustrate this
formalism by using numerical simulations.

\section{Theory}

In this section, we develop a mean-field method to evaluate the
magnetic induction \b{$\vec{B}$} produced by one neuron or by a
population of neurons, based on Maxwell theory of electromagnetism.

In a first step, we start from Maxwell equations in mean field
\cite{BedDes2011a} and in Fourier frequency space, to derive the
differential equation for the magnetic induction \b{$\vec{B}$}.  Note
that in principle, one should use the notation \b{$<\vec{B}>$} for the
spatial arithmetic average of \b{$\vec{B}$}, but in the rest of the
paper we will use the notation \b{$\vec{\textbf{B}}$} for simplicity.
The same convention will be used for the other quantities such as the
magnetic field \b{$\vec{\textbf{H}}$}, electric field
\b{$\vec{\textbf{E}}$}, electric displacement \b{$\vec{\textbf{D}}$},
electric potential \b{$\textbf{V}$}, magnetic vector potential
\b{$\vec{\textbf{A}} $}, free-charge current density
\b{$\vec{\textbf{j}}^{~f} $}, generalized current density
\b{$\vec{\textbf{j}}^{~g}$} \cite{BedDes2013} and the impedance of the
extracellular medium \b{$ \textbf{z}_{medium} $}. Note that taking the
spatial arithmetic average of the medium impedance implies to take the
harmonic mean over the medium admittance \b{\textbf{${ \gamma }$}},
because we have \b{$\textbf{z}_{medium}=\boldmath{1/\gamma
  }=\boldmath{ 1/(\sigma +i\omega\varepsilon }) $}.  

In a second step, we evaluate \b{$\vec{\textbf{B}}$} produced by a cylinder
compartment embedded in a complex extracellular medium.  We begin by
calculating the the boundary conditions of \b{$\vec{\textbf{B}}$} on the
surface of the cylinder compartment.  This method uses the same
approach results that we recently introduced and applied to calculate
the transmembrane electric potential in the same model
\cite{BedDes2013}.  This method will be used to calculate the boundary
conditions of \b{$\vec{\textbf{B}}$}, and these boundary conditions will then
be used to obtain an explicit solution of the differential equation
that \b{$\vec{\textbf{B}}$} must satisfy.  Next, we will explicitly calculate
the field \b{$\vec{\textbf{B}}$}.

In a third step, we use these results together with the superposition
principle to obtain a general method to calculate the field
\b{$\vec{\textbf{B}}$} produced by a large number of cylinder compartments,
which can be either define a single neuron dendritic morphology, or a
population of neurons.

\subsection{Differential equation for \b{$\vec{\textbf{B}}$}}
\label{sec2.1}

We now derive the differential equation for \b{$\vec{\textbf{B}}$} in mean
field and in an extracellular medium which is linear, heterogeneous
and scalar\footnote{\samepage Note that by definition, a given medium
  {\it linear} when the linking equations between the fields are
  convolution products that do not depend on the field intensities.  A
  medium is {\it scalar} when the parameters in the convolution
  products do not depend on direction in space (ie, are isotropic),
  which is a good approximation in a mean-field theory.}. In such
media, we consider the general case where there can be formation of
ions, through chemical reactions.  Such charge creation or
annihilation will determine additional current densities. At any time,
we have: 
\b{$$
\left \{
\begin{array}{ccccc}
\rho^{~c+}  + \rho^{~c-} &=&  0\\ \\
\vec{\textbf{j}}^{~c} &=& \vec{\textbf{j}}^{~+} + \vec{\textbf{j}}^{~-} 
&=& \rho^{c+} \vec{\textbf{v}}^{~+} + 
\rho^{c-} \vec{\textbf{v}}^{~-}
\end{array}
\right .
$$}
where \b{$\rho^{c+}$} and \b{$\rho^{c-}$} are the variations of
positive and negative charge densities, produced by chemical reactions
in a given volume.  These relations express the fact that the
free-charge density \corr{remains constant} when we have creation and
annihilation of ions, but that the non-conservation of the total
number of ions determines, in general, a current density of charge
creation \b{$\vec{\textbf{j}}^{~c}$} (because
\b{$\vec{\textbf{j}}^{~+}$} and \b{$\vec{\textbf{j}}^{~-}$}
necessarily have the same sign).

In such a case, according to classic electromagnetism theory, charge
densities and current densities are linked by two sets of equations.
The first set comprises four operatorial equations: \b{
\begin{equation}
\label{eq1}
\begin{array}{rclcrcclc}
\nabla\cdot\vec{\textbf{\textbf{D}}}~(\vec{x},\omega) &= & \rho^f~(\vec{x},\omega) &~(i)  & & \nabla\cdot\vec{\textbf{\textbf{B}}}~(\vec{x},\omega) &=& 0 &~(iii)\\\\
\nabla\times\vec{\textbf{E}}~(\vec{x},\omega) &=& -i\omega\vec{\textbf{B}}~(\vec{x},\omega) &~(ii)  & & \nabla\times\vec{\textbf{H}}~(\vec{x},\omega) &=& ~\vec{\textbf{j}}^{~g}~(\vec{x},\omega) + \vec{\textbf{j}}^{~c}~(\vec{x},\omega)  &~(iv)
\end{array}
\end{equation}}
Note that 
\b{$\vec{\textbf{j}}^{~g}= \vec{\textbf{j}}^{~f} + i\omega\vec{\textbf{D}}$} [Eq.~(\ref{eq1}~iv)].
where \b{$\vec{\textbf{j}}^{~f}$} is the free-charge current density and 
\b{$i\omega\vec{\textbf{D}}$} is the displacement current density.

A second set of equations comprises the two linking equations between
\b{$\vec{\textbf{D}}$} and \b{$\vec{\textbf{E}}$}, as well as \b{$\vec{\textbf{H}}$} and
\b{$\vec{\textbf{B}}$} interaction fields, and one linking equation between the
free-charge current density field \b{$\vec{\textbf{j}}^{~f}$} and
\b{$\vec{\textbf{E}}$}. Experiments \cite{Gabriel1996b,Cole} and theory
\cite{Landau} have shown that these linking equations can be
represented by the following convolution equations \b{
\begin{equation}
\label{eq2}
\left \{
\begin{array}{ccccccccc}
\vec{\textbf{D}}~(\vec{x},\omega) &=& \varepsilon~(\vec{x},\omega)~\vec{\textbf{E}}~(\vec{x},\omega) &~&(i) \\\\ 
\vec{\textbf{B}}~(\vec{x},\omega) &=& \mu~(\vec{x},\omega)~\vec{\textbf{H}}~(\vec{x},\omega) &~&(ii) \\\\
\vec{\textbf{j}}^f~(\vec{x},\omega) &=&  \sigma~(\vec{x},\omega)~\vec{\textbf{E}}~(\vec{x},\omega) &~&(iii)
\end{array}
\right .
\end{equation}}
for a linear and scalar medium.  Note that all of the above was
formulated in Fourier frequency space.

Assuming that if the base volume considered in the mean-field analysis
is large enough, we have at any time the same number of creation and
annihilation of ions, and we can write
\b{$\vec{\textbf{j}}^{~c}(\vec{x},t)\approx 0$}, so that the Fourier transform
of \b{$\vec{\textbf{j}}^{~c}(\vec{x},t)$} can be considered
zero for physiological frequencies\footnote{Note that it is clear that
  one can have fluctuations of the number of ions per unit volume,
  independently of the size considered, when the time interval is
  sufficiently small.  However, such contributions will necessarily
  participate to very high frequencies in the variation of
  \b{$\vec{\textbf{j}}^{~c}(\vec{x},\omega)$}), which are well outside the
  ``physiological'' range of measurable frequencies in experiments
  (about 1-1000~Hz).}.  This is equivalent to consider that the
current fluctuations caused by chemical reactions are negligible.  It
follows from Eqs.~(\ref{eq1}~iii) and (\ref{eq1}~iv): \b{
\begin{equation}
\label{eq3}
\nabla\times (\nabla\times\vec{\textbf{\textbf{B}}}) =
-\nabla^2\vec{\textbf{\textbf{B}}} + \nabla ( \nabla\cdot\vec{\textbf{\textbf{B}}})=
-\nabla^2\vec{\textbf{\textbf{B}}}
=\mu_o\nabla\times\vec{\textbf{\textbf{j}}}^{~g}.
\end{equation}
} where \b{$\vec{\textbf{\textbf{j}}}^{~g}$} is the generalized current density.  This
current can be expressed as \b{$\vec{\textbf{j}}^{~g}=\gamma
  ~\vec{\textbf{E}}=(\sigma+i\omega\varepsilon )~\vec{\textbf{E}}$}, where \b{$\gamma$}
is the admittance of the scalar medium (in mean-field\footnote{Note
  that in a mean-field theory, the electromagnetic parameters are
  calculated for a given volume, and therefore do not depend on
  spatial coordinates (for a sufficiently large volume).  However, the
  renormalization to obtain the ``macroscopic'' electric parameters
  results in a frequency-dependence of these parameters.  This occurs
  if electric parameters are not spatially uniform at microscopic
  scales, or from processes such as ionic diffusion, polarization,
  etc.  \cite{BedDes2011a,BedDes2006b,BedDes2009}.}; see also the linking
equations [Eqs. (\ref{eq2})]).  If the volume of the mean-field
formalism is large enough, the admittance does not depend on spatial
position, and we can write: \b{
\begin{equation}
\label{eq4}
\nabla\times\vec{\textbf{j}}^{~g}=
\gamma~\nabla\times\vec{\textbf{E}}=-i\omega\gamma \vec{\textbf{B}}
\end{equation}}
It follows that
\b{
\begin{equation}
\label{eq5}
\nabla^2\vec{\textbf{B}}=~i\omega\mu_o\gamma~ \vec{\textbf{B}} ~ .
\end{equation}}

Thus, one sees that in general, the differential equation for
\b{$\vec{\textbf{B}}$} depends on the admittance of the medium \b{$\gamma$}.
This is due to the fact that we have considered \b{$\nabla \times
  \vec{\textbf{j}}^g \neq 0$} in Eq.~(\ref{eq4}), which is equivalent to allow
electromagnetic induction to occur.  
%This is counter-intuitive, as
%\b{$\vec{\textbf{B}}$} is usually believed to be independent of the electric
%parameters of the medium.

We will see later that, for physiological frequencies, the righthand
term of Eq.~(\ref{eq5}) is negligible, so that we can in practice
calculate \b{$\vec{\textbf{B}}$} very accurately using the expression
\b{$\nabla^2\vec{\textbf{B}}=0$}. Note that this approximation amounts to
neglect electromagnetic induction effects in the context of natural
neurophysiological phenomena of low frequency (\b{$<1000~Hz$}) because
the righthand of Eq.~(\ref{eq5}) originates in the mathematic
formalization of electromagnetic induction (Faraday-Maxwell law,
Eq.~(\ref{eq1}~ii)).  However, it is important to keep in mind that the
righthand term in Eq.~(\ref{eq5}) cannot be neglected in the presence of
magnetic stimulation \cite{George1999}, because this technique uses
electromagnetic induction to induce currents in biological media.
Therefore, when considering magnetic stimulation, we will need to
update this formalism accordingly.

\subsection{Evaluation of \b{$\vec{\textbf{B}}$} }
\label{S2.2}
\begin{figure}[h!] 
\centering
\includegraphics[width=15cm]{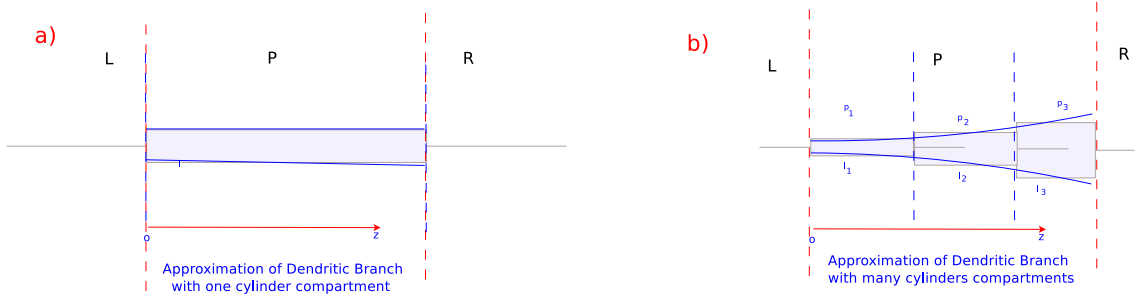}

\caption{\small(Color online) Scheme to calculate the magnetic
  induction produced by a dendritic branch. {\bf a.}  To evaluate the
  contribution of the dendritic segment, we divide space into three
  regions: L, P, R.  We first evaluate \b{$\textbf{B}^{\theta}$} in the
  principal region P, which corresponds to the space between Regions L
  and R.  Next, we evaluate \b{$\textbf{B}^{\theta}$} in the boundary regions
  \b{$L$} and \b{$R$}.  Note that the knowlegde of \b{$\textbf{B}^{\theta}$} in
  Region \b{$P$} is necessary to evaluate \b{$\textbf{B}^{\theta}$} in Regions
  \b{$L$} and \b{$R$} because it one must know \b{$\textbf{B}^{\theta}$} on the
  two planes \b{$z=0$} and \b{$z=\sum\limits_{i=1}^{N_p}l_i=l$}, in
  order to calculate its explicit value in Regions \b{$L$} and \b{$R$}
  using Eq.~(\ref{eq5}).  {\bf b.} Evaluation of \b{$\textbf{B}^{\theta}$} for
  a segment of variable diameter.  In this case, the same procedure is
  followed, except that Region~P is divided into $N_p$ compartments,
  each described by a continuous cylinder,
  \b{$P=\bigcup\limits_{i=1}^{N_p} p_i$}. Note that the continuity
  conditions on the axial current and the transmembrane voltage allow
  one to define boundary conditions for \b{$\textbf{B}^{\theta}$} over the
  surfaces of the compartments \b{$p_i$}. The figure shows an example
  with \b{$N_p=3$}.}

\label{fig1}
\end{figure}
In the preceding section, we have determined the differential equation
that \b{$\vec{\textbf{B}}$} must satisfy in Fourier frequency space.  Note that
the linearity of Eq.~(\ref{eq5}) implies that its solution for a given
frequency does not depend on other frequencies (which would not be
true if the equation was non-linear).  However, this equation is not
sufficient to determine \b{$\vec{\textbf{B}}$} because the boundary conditions
must be known to obtain an explicit solution.  To solve this boundary
condition problem, we must use cable equations because we consider the
``microscopic'' case where the electromagnetic field results from the
activity of each individual neuron, rather than considering
``macroscopic'' sources representing the activity of thousands of
neurons as traditionally done.  Moreover, to keep the formalism as
general as possible, we consider the ``generalized cable equations''
\cite{BedDes2013}, which generalizes the classic cable equations of
Rall \cite{Rall1962,Rall1995} to the general situation where the
extracellular medium can have complex or inhomogeneous electrical
properties.  We will also use a similar method of continuous cylinder
compartment as introduced previously \cite{BedDes2013}\footnote{The
  method of continuous cylinder compartments consists of solving
  analytically the cable equations in a continuous cylindric cable
  compartment, which can be of arbitrary length, but constant diameter
  (see details in \cite{BedDes2013}).}.

In the following, we first calculate the boundary conditions for an
arbitrary cylinder compartment (with arbitrary length and diameter)
\cite{BedDes2013}. We will see that it is sufficient to evaluate the
generalized axial current \b{$i_i^g$} inside each continuous cylinder
compartment to evaluate its boundary conditions.  Second, we consider
the more realistic scenarion of a dendritic branch of variable
diameter, which is approximated by continuous cylinder compartments
(Fig.~\ref{fig1}).  We then calculate everywhere in space the value of
\b{$\vec{\textbf{B}}$} produced by this dendritic branch.  Finally, we
give a general description of the computation of
\b{$\vec{\textbf{B}}$} produced by several dendritic branches.  This
description can apply in general to any dendritic morphology, or
axons, from one or several neurons.

\subsubsection{Boundary conditions of \b{$\vec{\textbf{B}}$} for a continuous
  cylinder compartment}

We now calculate the boundary conditions of \b{$\vec{\textbf{B}}$} on
the surface of a continuous cylinder compartment.  \corr{To do this,
  we set \b{$\vec{\textbf{B}}= \textbf{B}^{\theta}~\hat{e}^{\theta}$}
  because we have a complete cylindric symmetry (see details in
  Appendix~\ref{appA}).  Once the direction of \b{$\vec{\textbf{B}}$}
  is know, one can calculate the boundary conditions of
  \b{$\vec{\textbf{B}}$} using Amp\`ere-Maxwell's law}.

\begin{figure}[h!] 
\centering
\includegraphics[width=5cm]{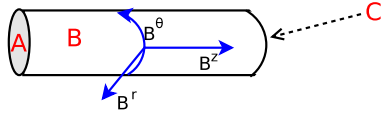}

\caption{\small (Color online) Coordinate ccheme for a cable segment
  of constant diameter. The scheme shows the cable with the cylindric
  coordinate system used in the paper, as well as the surfaces \b{$A$}
  and \b{$C$}, which are the sections that cuts the cable
  perpendicular to its membrane (delimited by surface \b{$B$}).
  \b{$\mathcal{D}$} is the interior volume of the segment, as
  delimited by these surfaces, and \b{$\partial \mathcal{D}$} is the
  reunion of the two surfaces \b{$A$} and \b{$B$}.}
 
\label{fig2}
\end{figure}

We now evaluate \b{$\textbf{B}^{\theta}$} as a function of the generalized current.  We
calculate the values of \b{$\textbf{B}^{\theta}$} as a function of the
generalized current over the surface \b{$\mathcal{S}_B$}
(Fig.~\ref{fig2}) using Amp\`ere-Maxwell law [Eq.~(\ref{eq1})~iv]. We
obtain: \b{
\begin{equation}
\label{eq6}
\oint\limits_{\partial\mathcal{S}_B} \vec{\textbf{B}}.d\vec{s} = 
\iint\limits_{\mathcal{S}_A}
\nabla\times\vec{\textbf{B}}\cdot\hat{n}_{\mathcal{S}_A}~dS =
\mu_o \iint\limits_{\mathcal{S}_A}\vec{\textbf{j}}^{~g}\cdot\hat{n}_{\mathcal{S}_A}~dS
 =\mu_o i_i^{~g} ~ ,
\end{equation}}
where \b{$i_i^{~g}$} is the generalized axial current inside the continuous
cylinder compartment.  Taking into account cylindic symmetry gives:
\b{
\begin{equation}
\label{eq7}
\vec{\textbf{B}} =\textbf{B}^{\theta}\hat{e}_{\theta}=\frac{\mu_o i_i^{~g} (z,\omega)}{2\pi a }~ \hat{e}_{\theta} ~ ,
\end{equation}}
where \b{$i_i^{~g}$} is the axial current inside the compartment and 
\b{$a$} is its radius.

This equation together with Eq.~(\ref{eq5}) show that the value of
\b{$\vec{\textbf{B}}$} around a dendritic compartment will depend on the
impedance of the extracellular medium (\b{$1/\gamma$}) for two
different reasons.  First, the righthand term of Eq.~(\ref{eq5})
explicitly depends on the extracellular impedance, but we will see in
the next section that these electromagnetic induction effects are
likely to be negligible.  Second, Eq.~(\ref{eq7}) shows that the
boundary conditions also depend on the extracellular impedance,
because the spatial and frequency profiles of \b{$i_i^{~g}$} depend on
this impedance \cite{BedDes2013}.  However, we will see that, contrary
to electromagnetic induction, this dependency cannot be neglected when
calculating \b{$\vec{\textbf{B}}$}, because this effect is potentially
important. In the next section, we calculate
magnetic induction in the extracellular space by directly solving
Eq.~(\ref{eq5}) using the boundary conditions evaluated by
Eq.~(\ref{eq7}).

\subsubsection{General expression of \b{$\vec{\textbf{B}}$} in extracellular
space for a dendritic branch \label{sec2.2.2}}

In this section, we derive a method to calculate the expression of
\b{$\vec{\textbf{B}}$} for a dendritic branch 
(Fig.~\ref{fig1})  In cylindric coordinates, Eq.~(\ref{eq5})
writes: \b{
\begin{equation}
%\begin{array}{ccc}
\begin{split}
\nabla^2\vec{\textbf{B}}  =  [\frac{\partial^2 \textbf{B}^r}{\partial r^2} + \frac{1}{ r^2}\frac{\partial^2 \textbf{B}^r}{\partial \theta^2} +
\frac{\partial^2 \textbf{B}^r}{\partial z^2} +
\frac{1}{ r}\frac{\partial \textbf{B}^r}{\partial r} 
-\frac{2}{ r^2}\frac{\partial \textbf{B}^{\theta}}{\partial \theta}
-\frac{\textbf{B}^r}{ r^2}] ~\mathbf{\hat{e}_r} \cdots\\
+ ~[\frac{\partial^2 \textbf{B}^{\theta}}{\partial r^2} + \frac{1}{ r^2}\frac{\partial^2 \textbf{B}^{\theta}}{\partial \theta^2} +
\frac{\partial^2 \textbf{B}^{\theta}}{\partial z^2} +
\frac{1}{ r}\frac{\partial \textbf{B}^{\theta}}{\partial r} 
+\frac{2}{ r^2}\frac{\partial \textbf{B}^{r}}{\partial \theta}
-\frac{\textbf{B}^{\theta}}{ r^2}]~\mathbf{ \hat{e}_{\theta} } \cdots\\
+~ [\frac{\partial^2 \textbf{B}^z}{\partial r^2} + \frac{1}{ r^2}\frac{\partial^2 \textbf{B}^z}{\partial \theta^2} +
\frac{\partial^2 \textbf{B}^z}{\partial z^2} +
\frac{1}{ r}\frac{\partial \textbf{B}^z}{\partial r}] ~\mathbf{ \hat{e}_{z} }
 = i\omega\mu_o\gamma ~\vec{\textbf{B}}  =   i\omega\mu_o\gamma~ [\textbf{B}^r\mathbf{ \hat{e}_{r} } + \textbf{B}^{\theta}\mathbf{ \hat{e}_{\theta} } + \textbf{B}^z   \mathbf{ \hat{e}_{z}] }  
 \end{split}
%\end{array}
\label{eq8}
\end{equation}}

According to preceding section, the boundary conditions imply
\b{$\vec{\textbf{B}} =\textbf{B}^{\theta} (r,z) ~\hat{e}_{\theta}$} on the surface of
each continuous cylinder compartment, as well as \b{$\vec{\textbf{B}}=0$} for
infinite distances.  The cylindric symmetry of the boundary conditions
implies that \b{$\textbf{B}^{r}=\textbf{B}^{z}=0$} everywhere in space because the
solution of Eq.~(\ref{eq8}) is unique.  Consequently, to evaluate the
value of \b{$\textbf{B}^{\theta}$} produced by a dendritic branch, one must
solve the following equation: \b{
\begin{equation}
\frac{\partial^2 \textbf{B}^{\theta}}{\partial r^2} +
\frac{1}{r}\frac{\partial \textbf{B}^{\theta}}{\partial r} +
\frac{\partial^2 \textbf{B}^{\theta}}{\partial z^2} 
-\frac{\textbf{B}^{\theta}}{r^2}
=i\omega\mu_o\gamma \textbf{B}^{\theta} ~ .
\label{eq9}
\end{equation}}

\subsubsection{Solving the equation of \b{$\vec{\textbf{B}}$} for a continuous
cylinder compartment}
\label{sec2.2.3}

\corr{In this section, we present an iterative method to calculate the
  solution of Eq.~(\ref{eq9}) in natural conditions (in the absence of
  electric or magnetic stimulation), and for a continuous cylinder
  compartment of radius \b{$a$} and length \b{$l$}, when the values of
  \b{$\textbf{B}^{\theta}$} on its surface are known.  To do this, we
  neglect electromagnetic induction and set the right term of
  Eq.~(\ref{eq9}) \b{$i\omega\mu_o\gamma \textbf{B}$} to zero,}
because we have \b{$\omega \mu_o |\gamma |\approx 0$} for the typical
size of a neuron in cerebral cortex, and for frequencies lower than
about \b{$1000~Hz$}. Indeed \b{$\mu_o = 4\pi\times10^{-7}~H/m$} and
the admittance of the extracellular medium is certainly lower than
that of sea water, and thus we can write \b{$|\gamma_{medium} |<
  |\gamma_{sea ~water} |<1$} \corr{and if we consider that
  \b{$r_{cortex}<<r_{max} = 1~m$}, then we have \b{$k^2 + 1/r^2 >
    1/r_{max}^2 = 1 >> \omega \mu_{0} |\gamma| $}}.  This
approximation amounts to neglect the phenomenon of electromagnetic
induction (in the absence of magnetic stimulation).  Thus, the
frequency dependence of \b{$\vec{\textbf{B}}$} is essentially caused
by the frequency dependence of the axial current \b{$i_i^g$}.  Note
that \b{$i_i^g$} depends on the nature of extracellular and cytoplasm
impedances, as shown previously \corr{in the generalized
  cable}~\cite{BedDes2013}.

\corr{The goal of this approach is to provide a method to solve
  Laplace's equation (\b{$\nabla^2\vec{\textbf{B}}=0$}) in 3D,
  assuming a perfect cylindric symmetry of the dendritic compartment.
This approach allows one to reduce the problem to two
  dimensions\footnote{Note that Laplace's equation can also be solved
    using the finite element method for a simple geometry.  For
    example, Galerkin \cite{Salon99} method works very well in this
    case, but requires significant computation time \corr{compared to
      a two-dimensional method}}.  We approach the solution of
  this problem by using an iterative method. The idea of the method is to
  calculate, in a first step, the solution using complex Fourier
  transform, which gives an exact solution for an infinite cylinder.
  This first estimate is then corrected by successive iterations using
  the first-order Hankel transform. This method is
  presented in detail in Appendix~\ref{appB}, while in
  Appendix~\ref{appC}, we demonstrate that the method converges.}

\subsubsection{The general expression of \b{$\vec{\textbf{B}}$} for \b{$N_B$}
  dendritic branches from one or several neurons}

Assuming that electromagnetic induction is negligible, and that the medium
is linear, we can apply the superposition principle such that we can write
\b{$\vec{\textbf{B}}$} as:
\b{
\begin{equation}
\vec{\textbf{B}} = \sum_{i=1}^{N_B}\vec{\textbf{B}}_{i} 
\label{eq10}
\end{equation}}
where each \b{$\vec{\textbf{B}}_{i}$} is the magnetic induction produced by each
branch as if it was isolated. 
\begin{figure}[h!] 
\centering
\includegraphics[width=8cm]{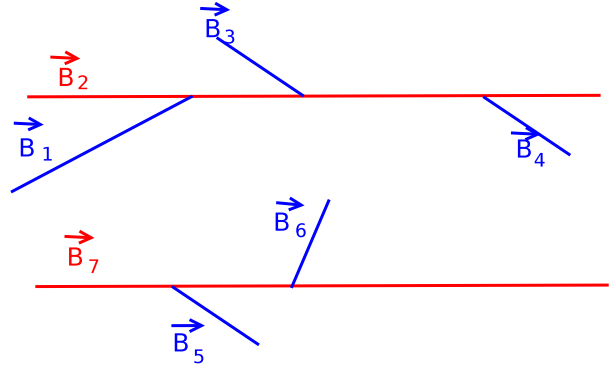}

\caption{\small (Color online) Example with 2 neurons.  In order to calculate the
  value of the magnetic induction \b{$\vec{\textbf{B}}$} generated by many
  neurons, one has to sum the values of \b{$\vec{\textbf{B}}_i$} produced by
  each branch.  Thus, it is sufficient to know the axial current
  \b{$i_i^{~g}$} at each branch to calculate \b{$\vec{\textbf{B}}$}.}
 
\label{fig3}
\end{figure}

Thus, at some distance away of an ensemble of dendritic branches
assimilable to continuous cylinder compartments (Fig. \ref{fig3}), the field
\b{$\vec{\textbf{B}} $} is the vectorial sum of the field \b{$\vec{\textbf{B}}$} produced
by each compartment, which is itself calculated from the average
spatial and frequency profile of the axial current in each compartment
(see Sec.~\ref{sec2.2.3}).

\subsection{Importance of the spatial profile of the axial current}
\label{sec2.3}

In the previous section, we have calculated \b{$\vec{\textbf{B}}$} without
explicitly considering the current in the extracellular space around
the neuron.  However, we know that this current necessarily produces a
magnetic induction, and thus it is necessary to include this
contribution to obtain a complete evaluation of \b{$\vec{\textbf{B}}$} in
extracellular space. In this section, we show that that this
contribution of extracellular currents is implicitly taken into
account by our formalism, through the spatial and frequency profile of
\b{$i_i^g$}.

According to Eqs.~(\ref{eq1}iv) and (\ref{eq2}ii), we can evaluate the
generalized current outside of a continuous cylinder compartment: \b{
\begin{equation}
        \vec{\textbf{j}}^{~g} = \frac{1}{\mu_o}\nabla\times\vec{\textbf{B}}
\label{eq11}
\end{equation}}
when \b{$\vec{\textbf{j}}^{~c} =0$} and for \b{$\mu(\vec{x},\omega)=\mu_o$}.  
Rewriting this expression in cylindric coordinates, we obtain
\b{
\begin{equation}
        \vec{\textbf{j}}^{~g} = \frac{1}{\mu_o}[~(\frac{1}{r}\frac{\partial \textbf{B}^z}{\partial\theta}-\frac{\partial \textbf{B}^{\theta}}{\partial z})~\hat{e}_r
        +
 (\frac{\partial \textbf{B}^r}{\partial z}-\frac{\partial \textbf{B}^{z}}{\partial r})~\hat{e}_{\theta}
 +
 \frac{1}{r}~(\frac{\partial (r\textbf{B}^{\theta})}{\partial r}-\frac{\partial \textbf{B}^{r}}{\partial \theta})~\hat{e}_z~]
        \label{eq12}
\end{equation}
}It follows that
\b{
\begin{equation}
        \vec{\textbf{j}}^{~g} = \frac{1}{\mu_o}[-\frac{\partial \textbf{B}^{\theta}}{\partial z}~\hat{e}_r
 +
(\frac{\partial \textbf{B}^{\theta}}{\partial r}+\frac{\textbf{B}^{\theta}}{r})~\hat{e}_z]
        \label{eq13}
\end{equation}}
because the solution is of the form \b{$\vec{\textbf{B}}(r,\theta,z,\omega) 
=\textbf{B}^{\theta}(r,z,\omega)~ \hat{e}_{\theta}$} [Sec.~\ref{sec2.2.2}]. 
We see that the generalized current density outside of the neuron is 
different from zero, if and only if we have
\b{
\begin{equation}
\left \{
\begin{array}{ccccccc}
-\frac{\partial \textbf{B}^{\theta}}{\partial z} &\neq &0
\\\\
\frac{\partial \textbf{B}^{\theta}}{\partial r}+\frac{\textbf{B}^{\theta}}{r} &\neq &0
\end{array}
\right .
\label{eq14}
\end{equation}}
Thus, the external current around the neuron is taken into account because the
solution depends on \b{$r$} and \b{$z$} in general (see preceding section).

\begin{figure}
\centering
\includegraphics[width=8cm]{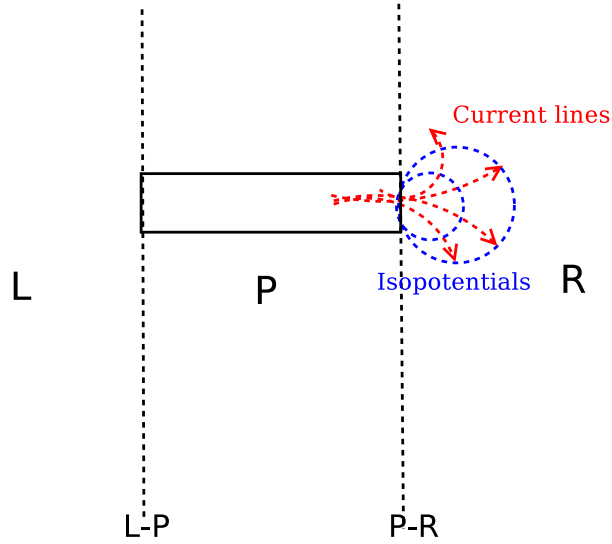}

\caption{\small (Color online)  Illustration of the
  current fields around the soma of a ball-and-stick model.  The
  current fields are shown (arrows) around the soma when the
  generalized membrane current is perpendicular to the soma membrane
  (red arrows).  The isopotential surfaces are shown in blue and
  correspond to the soma membrane .  If the soma has a different
  ``diameter'', but coincides with the isopotential surface, then the
  geometry of these current lines and isopotential surfaces remains
  invariant.  However, the value of the electric potential is
  different on each equipotential surface.}

\label{fig4}
\end{figure}

Note that we have \b{$\frac{\partial \textbf{B}^{\theta}}{\partial z} =0$}
(Fig.~\ref{fig1}) if and only if the spatial profile of the axial
current \b{$i_i^g$} does not depend on \b{$z$} [Eq.~(\ref{eq7})].  In
this case, the current \b{$i_m$} is zero, which implies that the
electric field produced by the compartment is also zero
\cite{Rall1962,Rall1995,BedDes2013}. In addition, we know that in a
neuron, one cannot have axial current without transmembrane current,
and thus, it is impossible that \b{$\frac{\partial
    \textbf{B}^{\theta}}{\partial z} =0$} in a given compartment.  Therefore,
we can conclude that the external current is taken into consideration
because \b{$\nabla\times\vec{\textbf{B}}\neq 0$} outside of the compartment
when \b{$\vec{\textbf{B}}$} depends on \b{$z$}.

In the preceding section, we have calculated \b{$\textbf{B}^{\theta}$}
for a single continuous cylinder compartment.  We now consider the
more complex case when this compartment is connected to a soma on one
side, according to a ``ball-and-stick'' configuration.  In this case,
one can consider that the current density \b{$\vec{\textbf{j}}^{~g}$}
in Region~R satisfies \b{$\nabla\cdot\vec{\textbf{j}}^{~g}=0$}
(generalized current conservation law) when
\b{$\vec{\textbf{j}}^{~c}=0$} and
\b{$$\nabla\times\vec{\textbf{j}}^{~g}= (\sigma_e+i\omega\varepsilon
  )~ \nabla\times\vec{\textbf{E}}=0$$} (when electromagnetic induction
is negligible, and in mean-field)\footnote{Note that we have
  considered several scales in \cite{BedDes2013}: the interior of the
  dendritic compartment, the interior of the soma, the membrane, and
  the extracellular medium.}.  It follows that we have
\b{$\nabla^2\vec{\textbf{j}}^{~g}=0$} in each point of Region~R. Thus,
the field \b{$\vec{\textbf{j}}^{~g}$} does not explicitly depend on
electomagnetic parameters.  With the continuity condition of the
current at the interface between Regions~P and R, and the vanishing at
infinite distances (\b{$\vec{\textbf{j}}^{~g}
  \overset{\infty}{\longrightarrow} 0$}), we have a unique solution
(Dirichlet problem) in Region~R (Fig. \ref{fig4}).

However, the method to calculate the generalized cable for the
ball-and-stick model implicitly considers the soma impedance in the
spatial and frequency profiles on the continuous cylinder
compartment(s) ~\cite{BedDes2013}\footnote{In this paper, we have
  assumed that \b{$\vec{\textbf{j}}^{~g}$} is perpendicular to the membrane
  surface at the soma.  This implies that the internal and external
  surfaces of the soma are equipotential because
  \b{$(\sigma_e+i\omega\varepsilon)\vec{\textbf{E}}$} is perpendicular to the
  soma membrane.  Thus, the soma membrane is characterized by an
  impedance \b{$Z_s = \frac{V_m}{i_i^g}$}, which affects the spatial
  and frequency profiles in the dendritic compartments.}.  Thus, the
soma impedance is also taken into account implicitly here when
calculating the current at the interface between Regions P and R.

It is important to note that the same current geometries can be seen
for different soma sizes (Fig.~\ref{fig4}), and thus different neuron
models of identical dendritic structure but different soma will
generate identical magnetic inductions in Region~R (comprising the
soma).  Note that it does not apply to the electric field and
potential around the soma because we have
\b{$\vec{\textbf{E}}=\frac{\vec{\textbf{j}}^{~g}}{(\sigma_e+i\omega)\varepsilon}$}
where \b{$(\sigma_e+i\omega\varepsilon)$} depends on the size of the
soma membrane.  Thus, the soma impedance is sufficient to determine
\b{$\vec{\textbf{B}}$} but its exact size is not important if the soma
coincides with an isopotential surface.

Consequently, taking into account the spatial and frequency profiles
of \b{$\textbf{B}^{\theta}$} over the surface of the cylinder compartments
allows one to calculate everywhere in space the field \b{$\vec{\textbf{B}}$} as
well as the current fields inside and outside of the membrane.  Thus,
the spatial and frequency profiles of \b{$i_i^g$} [Eq. (\ref{eq7})]
implicitly take into account the screening effect caused by the
``return current'' outside of the neuron, when present.  Note that
this conclusion is entirely consistent with Maxwell equations and the
pseudo-parabolic equation (\ref{eq9}) derived from it, because these
equations determine a unique solution for a given set of boundary
conditions.  In the next section, we show how this method can be
generalized to complex morphologies or populations of neurons (still
under the condition that electromagnetic induction is negligible).

\section{Numerical simulations}\label{S3}

In this section, we show a few simulations with different types of
media for a ball-and-stick type model.  In a first step, we describe
how to calculate the generalized axial current as a function of the
synaptic current for a ball-and-stick type model.  In a second step,
we apply the method developed above to calculate the magnetic
induction.  We show here two examples, first when the extracellular
and cytoplasm impedances are resistive, and second, when these two
impedances are diffusive (Warburg impedance).

\subsection{Method to calculate the generalized axial current for a
  ball-and-stick model}\label{S3.1}

\begin{figure} 
\centering
  \includegraphics[width=8cm]{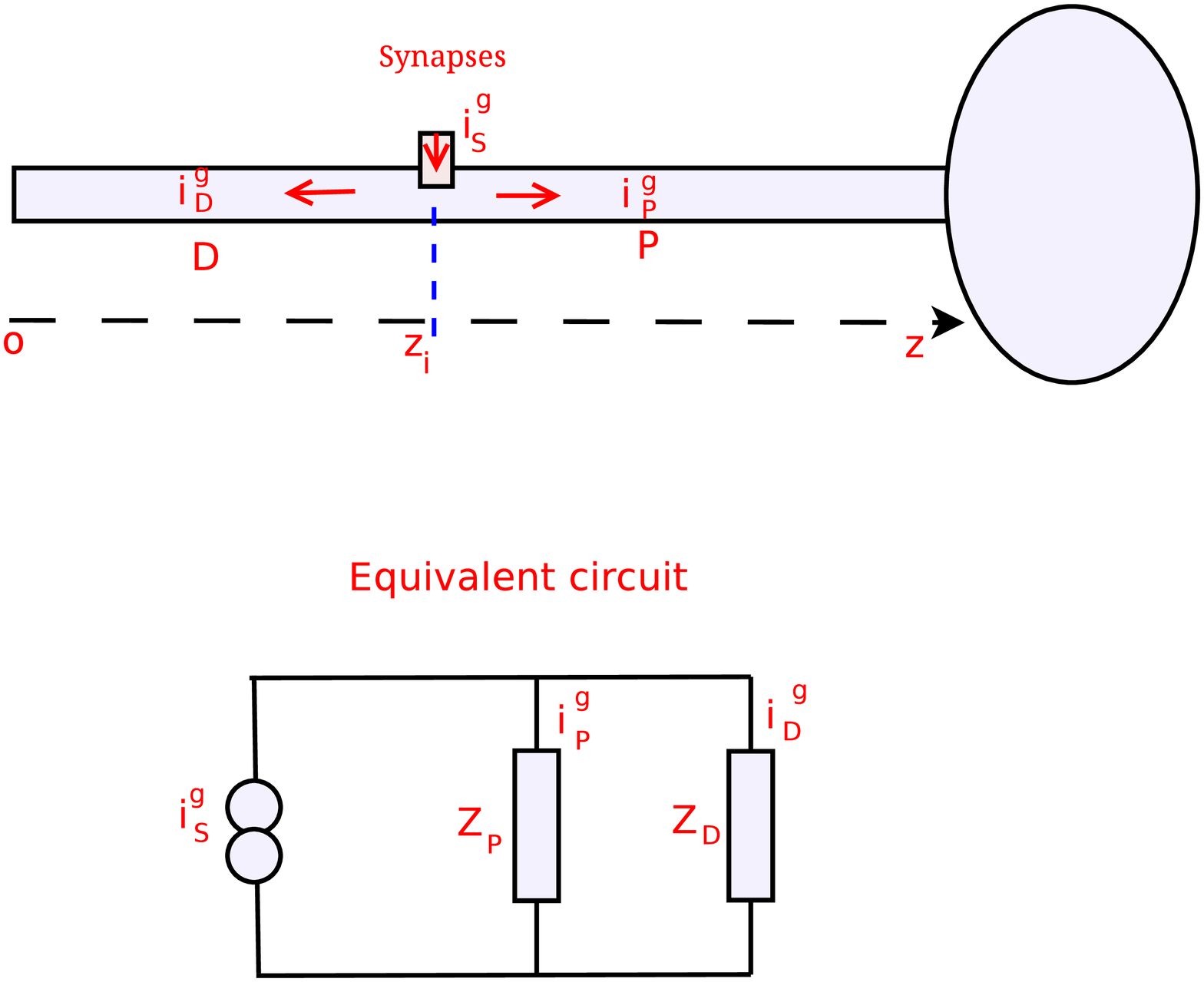}
  \caption{\small (Color online) Equivalent scheme to calculate the current flowing from
    distal to proximal at the position of the synapse, when the
    synaptic current is known. }
\label{fig5}
\end{figure}

In a first step, we determine the transmembrane voltage in the
postsynaptic region.  The current produced in this region separates in
two parts: one that goes to the soma (``proximal''), and another one
going in the opposite direction (``distal'') (Fig.~\ref{fig5}).
These two currents are given by the following relations,
\b{$Z_D(z_i,\omega)=\frac{V_m(z_i,\omega)}{i_{iD}^{~g}(z_i,\omega)}$}
and
\b{$Z_P(z_i,\omega)=\frac{V_m(z_i,\omega)}{i_{iP}^{~g}(z_i,\omega)}$},
for the distal and proximal  regions, respectively.  These expressions
were derived previously~\cite{BedDes2013}.  

Next, we determine the equivalent impedance at the position of the
synapse (Fig.~\ref{fig5}) \cite{BedDes2013}. We obtain \b{
\begin{equation}
Z_{eq}(z_i,\omega) = \frac{Z_P(z_i,\omega)Z_D(z_i,\omega)}{Z_P(z_i,\omega)+Z_D(z_i,\omega)}
\label{eq15}
\end{equation}}

It follows that the transmembrane voltage at the
position of the synapse is given by: \b{
\begin{equation}
V_m(z_i,\omega) = Z_{eq}(z_i,\omega)~i_s^{~g}(z_i,\omega)
\label{eq16}
\end{equation}}
when the synapse is at position \b{$z_i$}.  Next, we determine 
\b{$i_A^g(z_i,\omega)$} and \b{$i_D^g(z_i,\omega)$} from the 
following expressions: 
\b{
\begin{equation}
\left \{
\begin{array}{cccc}
i_P^{~g}(z_i,\omega) & = & \frac{V_m(z_i,\omega)}{Z_P(z_i,\omega)} \\\\
i_D^{~g}(z_i,\omega) & = & \frac{V_m(z_i,\omega)}{Z_D(z_i,\omega)}
\end{array}
\right  .
\label{eq17}
\end{equation}}

We have seen in \cite{BedDes2013} that with the generalized current,
the cable equations can be written in a form similar to the standard
cable equation: \b{
\begin{equation}
      \frac{\partial^2 V_m(z,\omega)}{\partial z^2} = \kappa_{\lambda}^2~ V_m(z,\omega)
\label{eq18}
\end{equation}}
where
\b{
\begin{equation}
\begin{array}{ccc}
 \kappa_{\lambda}^2 & = & \frac{\bar{z}_i~(1+i\omega\tau_m)}{r_m}
 = \frac{z_i~(1+i\omega\tau_m)}{r_m~[1+\frac{z_e^{(m)}}{r_m}(1+ i\omega \tau_m)]} 
\end{array} ~ , 
\label{eq19}
\end{equation}}
where \b{$1/r_m$}, \b{ $z_i$} and
\b{$\tau_m$} are, respectively, the linear density of
membrane conductance (in \b{$S/$m}), the impedance per unit length of the
cytoplasm (in
[\b{$\Omega / $m}]) and the membrane time constant. The
parameter \b{$z_e^{(m)}$} stands for the specific impedance of the extracellular
medium.  This parameter impacts on the spatial and frequency profile  
of \b{$V_m$}, \b{$i_m$} and \b{$i_i^{~g}$}, and has the same units as 
\b{$r_m$}.

The general solution of this equation in Fourier space \b{$\omega \neq
  0$} is given by: \b{
\begin{equation}
\left \{
\begin{array}{cccccc}
 V_{mD}(z,\omega) &=& A_P^{+}(z_i,\omega)~e^{+\kappa_{\lambda} z}
&+& A_D^{-}(z_i,\omega)~e^{-\kappa_{\lambda} 
z}\\\\
V_{mP}(z,\omega) &=& A_P^{+}(z_i,\omega)~e^{+\kappa_{\lambda} (l-z)}
&+& A_P^{-}(z_i,\omega)~e^{-\kappa_{\lambda}(l-z)}
\end{array}
\right .
\label{eq20}
\end{equation}}
for a continuous cylinder compartment of length \b{$l$} and constant
diameter, and when we know the synaptic current at position \b{$z=z_i$}. 
In such conditions, the coefficients of Eq.~(\ref{eq15}) are given by
the following expressions (see Appendix~F in \cite{BedDes2013}):
\b{
\begin{equation}
\begin{array} {ccccc}
~&
\left \{ 
\begin{array}{ccccccc}
A_D^{+}(z_i,\omega) &=& \frac{1}{2}e^{-\kappa_{\lambda}z_i}~[~V_{mD}(z_i,\omega) +\frac{\bar{z}_i}{\kappa_{\lambda}}~i_{iD}^{~g}(z_i,\omega)~]  \\\\
A_D^{-}(z_i,\omega) &=& \frac{1}{2}e^{+\kappa_{\lambda}z_i}~[~V_{mD}(z_i,\omega) -\frac{\bar{z}_i}{\kappa_{\lambda}}~i_{iD}^{~g}(z_i,\omega)~]
\end{array}
\right .
\\\\\\
~&
\left \{ 
\begin{array} {ccccccc}
A_P^{+}(z_i,\omega) &=& \frac{1}{2}e^{-\kappa_{\lambda}(l-z_i)}~[~V_{mP}(z_i,\omega) +\frac{\bar{z}_i}{\kappa_{\lambda}}~i_{iP}^{~g}(z_i,\omega)~]  \\\\
A_P^{-}(z_i,\omega) &=& \frac{1}{2}e^{+\kappa_{\lambda}(l-z_i)}~[~V_{mP}(z_i,\omega) -\frac{\bar{z}_i}{\kappa_{\lambda}}~i_{iP}^{~g}(z_i,\omega)~]    
\end{array}
\right .
\end{array}
\label{eq21}
\end{equation}}
Note that we can verify that \b{$V_m$} is continuous, in which case we
have \b{$V_{mP}(z_i,\omega)=V_{mD}(z_i,\omega)$}, which is consistent with
the fact that the electric field is finite.  Thus, one sees that when
the synaptic current is known at a given position, the spatial profile
of \b{$V_m$} can be calculated exactly for a continuous cylinder
compartment.

It follows that one can deduce the spatial and frequency profiles of
\b{$V_m$} when we know the current generated by each synapse, thanks
to the superposition principle.
Finally, one can directly calculate the generalized current
by applying the following equation :
\b{
\begin{equation}
i_i^{~g} = - \frac{1}{\bar{z}_i}\frac{\partial V_m}{\partial z}
\label{eq22}
\end{equation}}
on Eq.~(\ref{eq10}) \cite{BedDes2013} .  We obtain the generalized axial current generated
by a single synapse: 
\b{
\begin{equation}
\left \{
\begin{array}{cccccc}
 i_{iD}^{~g}(z,\omega) &=& -\frac{\kappa_{\lambda}}{\bar{z}_i}~[~A_D^{+}(z_i,\omega)~e^{+\kappa_{\lambda} z}
&+& A_D^{-}(z_i,\omega)~e^{-\kappa_{\lambda} 
z}~]\\\\
i_{iP}^{~g}(z,\omega) &=& +\frac{\kappa_{\lambda}}{\bar{z}_i}~[~A_P^{+}(z_i,\omega)~e^{+\kappa_{\lambda} (l-z)}
&-& \frac{\kappa_{\lambda}}{\bar{z}_i}A_P^{-}(z_i,\omega)~e^{-\kappa_{\lambda}(l-z)}~]
\end{array}
\right .
\label{eq23}
\end{equation}}

To obtain the total axial current, one has just to sum up the
contributions of each synapse. Note that this ``linear'' 
assumption only holds for current-based inputs, and a modified model
is needed to account for conductance-based inputs (not shown).

Finally, the knowledge of the generalized axial current permits to
determine the boundary conditions on \b{$\vec{\textbf{B}}$} and apply
the method developed above [Eq.~(\ref{eq7})].  In the next section,
we apply this strategy to calculate the magnetic induction in
different situations.

\subsection{Simulations of \b{$\vec{\textbf{B}}$} in extracellular
  space}\label{S3.2}

In this section, we apply the theory to a ball-and-stick type model of
the neuron \cite{Tuckwell,Wu}, using two different approximations of the extracellular
medium and cytoplasm impedance, either when they are purely resistive
(Ohmic), or when ionic diffusion is taken into account, resulting in
Warburg type impedances \cite{BedDes2013}.

\begin{figure}[bht]
\centering
\includegraphics[width=14cm]{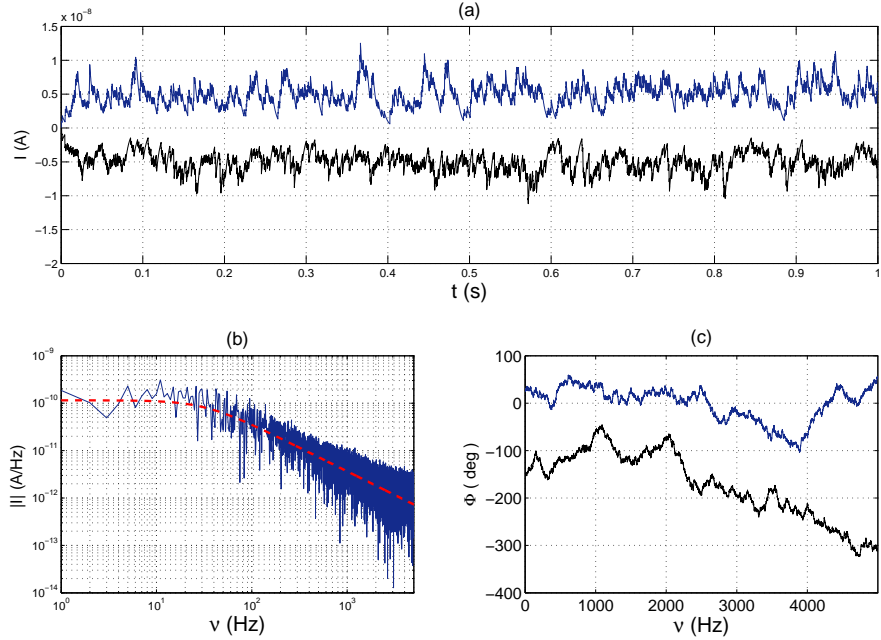}

\caption{\small (Color online) Synaptic current sources used in the
  simulations.  (a) Example of excitatory (blue, top curve) and
  inhibitory (black, bottom curve) current sources used in
  simulations.  These examples consists of 1000 random synaptic events
  per second.  (b) and (c): Modulus and phase, respectively, of the
  complex Fourier transform of these processes.  Note that the
  inhibitory current in not represented in (b) because its modulus is
  identical to that of the excitatory current.  The red dashed line in
  (b) corresponds to a Lorentzian (\b{$\frac{A}{1+i\omega\tau_m}$})
  with \b{$\tau_m = 5~ms$} and \b{$|A|=1~nA$}).}

\label{fig6}
\end{figure}

To do this, we model the ensemble of synaptic current sources as a
``stochastic dipole'' consisting of two stochastic currents, stemming
from excitatory and inhibitory synapses.  Each synaptic current is
described by a shot-noise given by:
\b{\begin{equation}
i_s = \sum_{n=1}^N c H(t-t_n)~e^{-(t-t_n)/\tau_m}
\label{eq24}
\end{equation}}
where \b{$H$} is the Heaviside function.  The stochastic variable
\b{$t_n$} follows a time-independent law.  We have chosen \b{$\tau_m =
  5~ms$} which corresponds to \textit{in vivo} conditions, \b{$c=
  +1~nA$} for excitatory synapses, and \b{$c= -1~nA$} for inhibitory
synapses (Fig.~\ref{fig6}).

In the simulations, we have simulated a ball-and-stick neuron model
with a dendrite of \b{$600~\mu m$} length and \b{$2~\mu m$} constant
diameter, and a spherical soma of \b{$7.5~\mu m$} radius.  The
synaptic currents were located at a distance of \b{$57.5~\mu m $} of
the soma for inhibitory synapses, and respectively \b{$357.5~\mu m $}
for excitatory synapses. Note that this particular choice was made
here to simplify the model.  This arrangement generates
a dipole which approximates the fact that inhibitory synapses are more
dense in the soma/proximal region of the neuron, while excitatory
synapses are denser in more distal dendrites \cite{DeFelipe}.

\subsubsection{Magnetic induction generated by a ball-and-stick model with resistive 
media
\label{sec3.2.1} }

We start by calculating the magnetic induction for the ``standard
model'' where the extracellular medium and cytoplasm are both
resistive.  The electric conductivity of cytoplasm was of \b{$3~S/m$},
and that of the extracellular medium was of \b{$5~S/m$}, in agreement
with previous models \cite{Rall1962,Rall1995,Wu,Koch}.

\begin{figure}[bht!]
\centering
  \includegraphics[width=14cm]{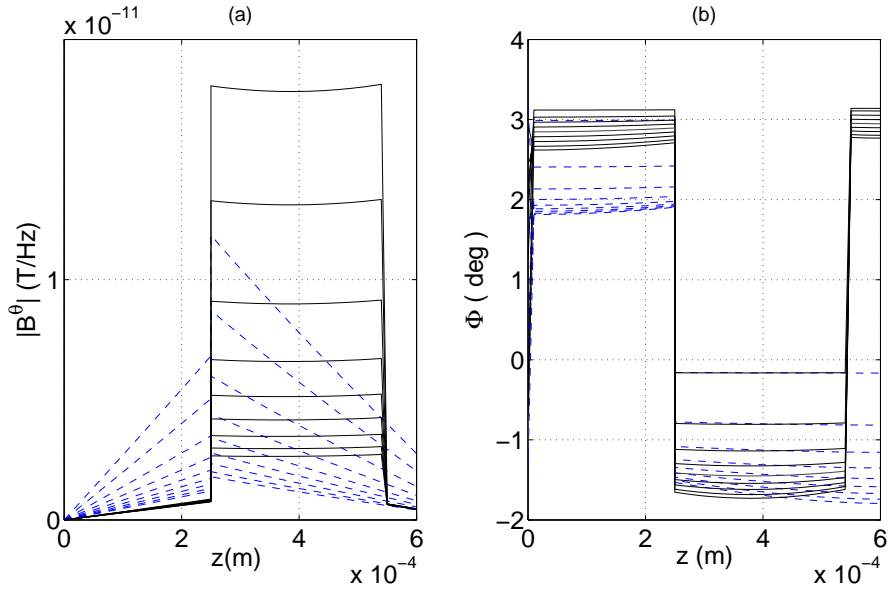}

  \caption{(Color online) Magnetic induction for the resistive model.
    \b{$\textbf{B}^{\theta}$} is shown here at the surface of the
    dendrite, as a function of position (distance to soma)
    for different frequencies between 1~Hz and
    \b{$5000~Hz$}. The blue dashed lines correspond to
    \b{$\textbf{B}^{\theta}$} generated when only excitatory synapses
    were present, and the black curves correspond to both synapses
    present.  \b{$\textbf{B}^{\theta}$} is always decreasing with
    frequency, and is larger and approximately constant between the
    two locations of the synaptic currents. }

\label{fig7}
\end{figure}

The magnetic induction generated by the resistive model is described
in Fig.~\ref{fig7}.  We can see that, for a given frequency, the
modulus of \b{$\textbf{B}^{\theta}$} is almost constant in space over
the dendritic branch in the region between the two locations of the
synaptic currents.  It is also smaller outside of this region.  Note
that the attenuation of \b{$\textbf{B}^{\theta}$} is completely
different whether excitatory or inhibitory synapses are present
(Fig.~\ref{fig7}, blue dashed curves).  Finally, we also see that the
attenuation of the axial current is very close to a linear law
although in reality we have a linear combination of exponentials (see
Eq.~\ref{eq24}).

\begin{figure}[bht!]
\centering
\includegraphics[width=14cm]{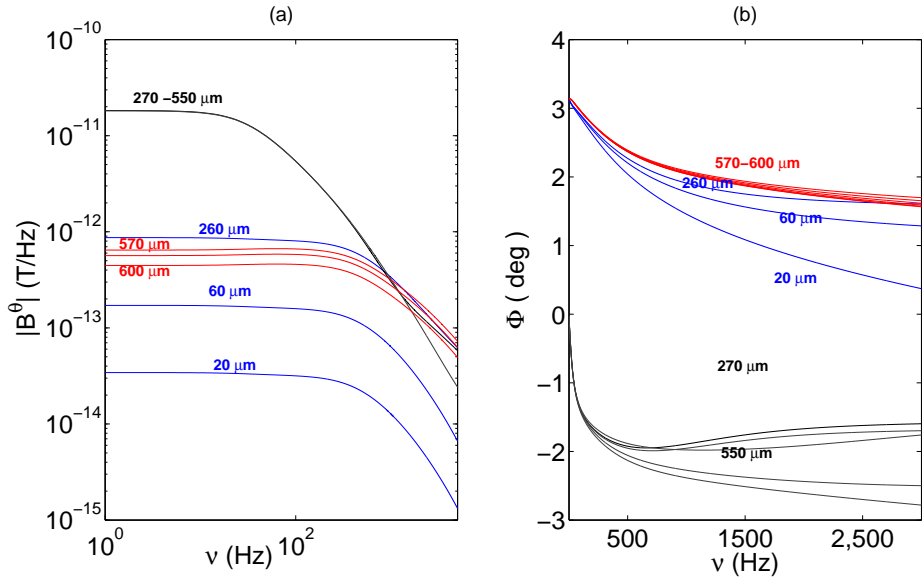}

\caption{(Color online) Frequency profile of the magnetic field for the resistive
  model.  \b{$\textbf{B}^{\theta}$} at the surface of the dendrite is
  represented as a function of frequency at different positions (both
  excitatory and inhibitory synapses were present).  The red curves
  correspond to different positions between the inhibitory synapses
  and the soma, the blue curves are taken at different positions
  between the excitatory synapses and the end of the dendrite, and the
  black curves represent positions in between the two synapse sites.
  Note that the modulus of \b{$\textbf{B}^{\theta}$} does not depend
  on position.}

\label{fig8}
\end{figure}

The frequency dependence of \b{$\textbf{B}^{\theta}$} is shown in
Fig.~\ref{fig8} for the resistive model.  The frequency dependence
depends on the position on the dendrite.  Between the two synapse
sites (black curves), the frequency dependence does not depend on the
position, and the scaling exponent is close to -1.5.  However, the
phase of \b{$\textbf{B}^{\theta}$} is position dependent, but is very
small (between 0 and -3 degrees).  In this region, the frequency
scaling begins at frequencies larger than about 10~Hz.

In the ``proximal'' region, between the soma and the location of
inhibitory synapses, the frequency dependence is different according
to the exact position on the dendrite (Fig.~\ref{fig8}, red curves)
and the frequency scaling occurs at frequencies larger than 1000~Hz.
However, the frequency scaling is almost identical and the exponent is
of about -1. The contribution of this region to the value of
\b{$\textbf{B}^{\theta}$} can be negligible compared to the preceding
region for the frequency range considered here ($<$1000~Hz). The phase
also shows little variations and is of small amplitude (between 1 and
3 degrees).

Finally, for the ``distal'' region, away of the site of excitatory
synapses, the frequency-dependence of the modulus of
\b{$\textbf{B}^{\theta}$} varies with the position on the dendrite,
and is significant only from about 1000~Hz, similar to the proximal
region.  The dependencies are almost identical between proximal and
distal regions, except for frequencies larger than 1000~Hz.  Note that
the contribution of these two regions to the value of
\b{$\textbf{B}^{\theta}$} is very small and can be considered
negligible compared to the region between the two synaptic sites (for
frequencies smaller than 1000~Hz).  The Fourier phase shows little
variations between 1 and 5000~Hz.  The frequency scaling exponent is
of the order of -1.5 between 2000 and 4000~Hz.  Note that the
numerical simulations also indicate that the boundary conditions on
the stick are very sensitive to the cytoplasm resistance but are less
sensitive to the extracellular resistance.

\subsubsection{Magnetic induction generated by a ball-and-stick model
  with diffusive media}\label{S3.2.2}

\begin{figure}[bht]
\centering
  \includegraphics[width=12cm]{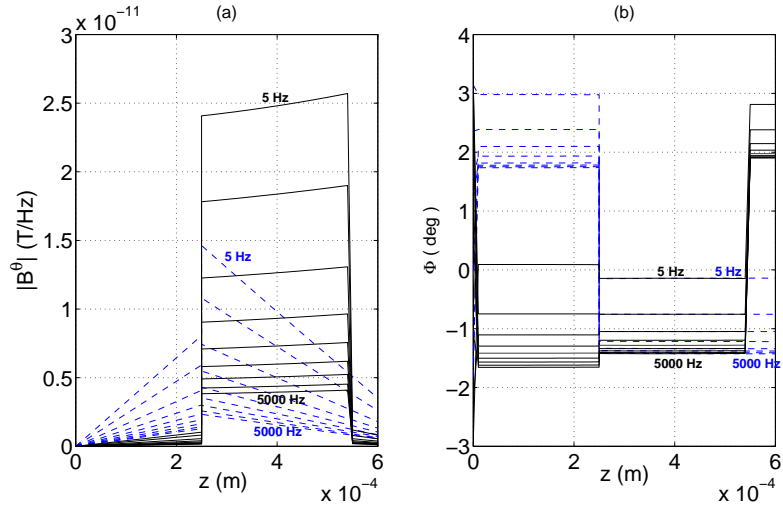} 

  \caption{\small (Color online) Magnetic induction
    \b{$\textbf{B}^{\theta}$} on the surface of the dendrite for a
    neuron embedded in diffusive media. \b{$\textbf{B}^{\theta}$} is
    represented for different frequencies. The blue dashed curves
    correspond to \b{$\textbf{B}^{\theta}$} produced at the surface of
    the dendrite with only excitatory synapses, and black curves
    correspond to excitatory and inhibitory synapses present.  We see
    that \b{$\textbf{B}^{\theta}$} is a decreasing function of
    frequency, and is higher towards inhibitory synapses, and low
    outside of this region.}

   \label{fig9}
\end{figure}

\begin{figure}[bht]
\centering
\includegraphics[width=12cm]{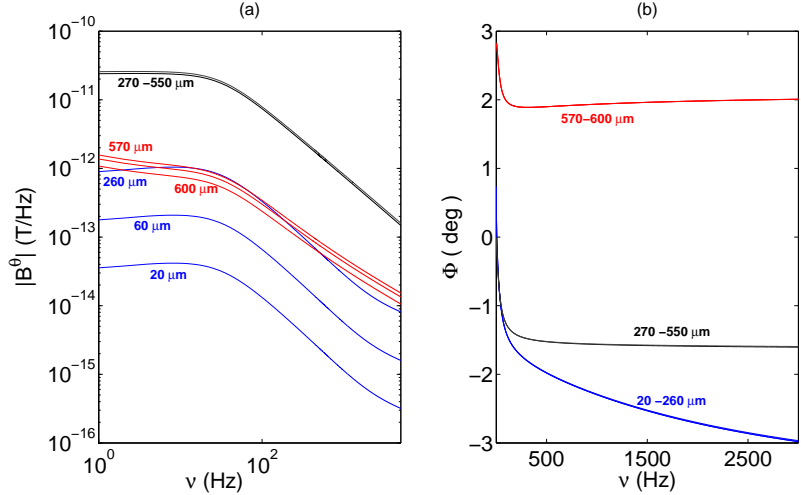}

\caption{\small (Color online) Magnetic induction
  \b{$\textbf{B}^{\theta}$} on the surface of the dendrite, as a
  function of frequency, for a neuron within diffusive media.
  \b{$\textbf{B}^{\theta}$} is represented for different positions on
  the dendrite, with both excitatory and inhibitory synapses present.
  One can see three distinct regions: proximal region between the soma
  and the location of inhibitory synapses (red curves), region between
  the two synaptic sites (black curves), and the distal region between
  the location of excitatory synapses and the end of the dendrite
  (blue curves).  Note that the modulus of \b{$\textbf{B}^{\theta}$}
  depends very weakly on dendritic position when we are in between the
  two synaptic sites.}

    \label{fig10}
\end{figure}

We now illustrate the same example as above, but when the
intracellular (cytoplasm) and extracellular media are described by a
diffusive-type Warburg impedance (Figs.~\ref{fig9} and \ref{fig10}).
We have assumed that the cytoplasm admittance is
\b{$\gamma=3\frac{\sqrt{\omega}~(1+i)}{\sqrt{2}}~S/m$}, while that of
the extracellular medium is
\b{$5\frac{\sqrt{\omega}~(1+i)}{\sqrt{2}}~S/m$}.  These values were
chosen such that the modulus of the admittance is the same as the
preceding example with resistive media (see Section~\ref{sec3.2.1})
for \b{$\omega=1~Hz$}.

When calculating the magnetic induction, we see that the modulus of
\b{$\textbf{B}^{\theta}$} on the dendrite surface increases when one
approaches the position of inhibitory synapses, but is very small
outside of this region (Fig.~\ref{fig9}, black curves).  Note that
the attenuation law of \b{$\textbf{B}^{\theta}$} along the dendritic
branch is completely different from that with only excitatory 
synapses present (Fig.~\ref{fig9}, blue dashed curves).  We also
see that the attenuation of the axial current is very close to a 
straight line, but in reality it is given by a sum of exponentials
(see Eqs.~\ref{eq24}).  

We can also see that the frequency dependence of
\b{$\textbf{B}^{\theta}$} depends on the region considered in the
dendrite (Fig.~\ref{fig10}).  In between the two synaptic sites (black
curves in Fig.~\ref{fig10}), the frequency dependence is almost
indepenent of position, with a scaling exponent close to -1 (in the
resistive case, it was -1.5 for the same conditions; see
Fig.~\ref{fig8}).  The Fourier phase of \b{$\textbf{B}^{\theta}$}
displays little variation.  The frequency dependence begins at a
frequency around 30~Hz.

In the ``proximal'' region, from the soma to the beginning of the
dendrite, the frequency dependence of the modulus of
\b{$\textbf{B}^{\theta}$} depends on position, and is present at all
frequency bands.  Between 1 and 10~Hz, the scaling exponent is close
to $1/4$, which would imply a PSD proportional to \b{$1/f^{1/2}$}.
This result is very different from the resistive case, which had a
negligible dependence at those frequencies (see Fig.~\ref{fig8}).
Note that the contribution of this region to the value of
\b{$\textbf{B}^{\theta}$} can be considered negligible compared to
the preceding region, for all frequencies between 1 and 5000~Hz
(which was not the case for resistive media; see Fig.~\ref{fig8}).
Finally, the Fourier phase is positive and approximately constant
for those frequencies.  The scaling exponent is -0.5 between 2000 
and 4000~Hz, while it was -1 in the resistive case examined above.

Finally, for the ``distal'' region, at the end of the dendrite, the
frequency dependence of the modulus of \b{$\textbf{B}^{\theta}$}
varies with position, and we observe a resonance around 30~Hz
(Fig.~\ref{fig10}).  A similar resonance was also seen previously in
the cable equation for diffusive media \cite{BedDes2013}. Similar to
the proximal region, the contribution of the distal region to the
value of \b{$\textbf{B}^{\theta}$} is very weak (for frequencies lower
than 1000~Hz).  The Fourier phase shows little variations.  The
scaling exponent is around -1 betwen 2000 and 4000~Hz, similarly to
the region between the synaptic sites.  As above, the boundary
conditions of the surface of the ``stick'' are much more sensitive to
the cytoplasm impedance.

\begin{figure}[bht!]
\centering
\includegraphics[width=8cm]{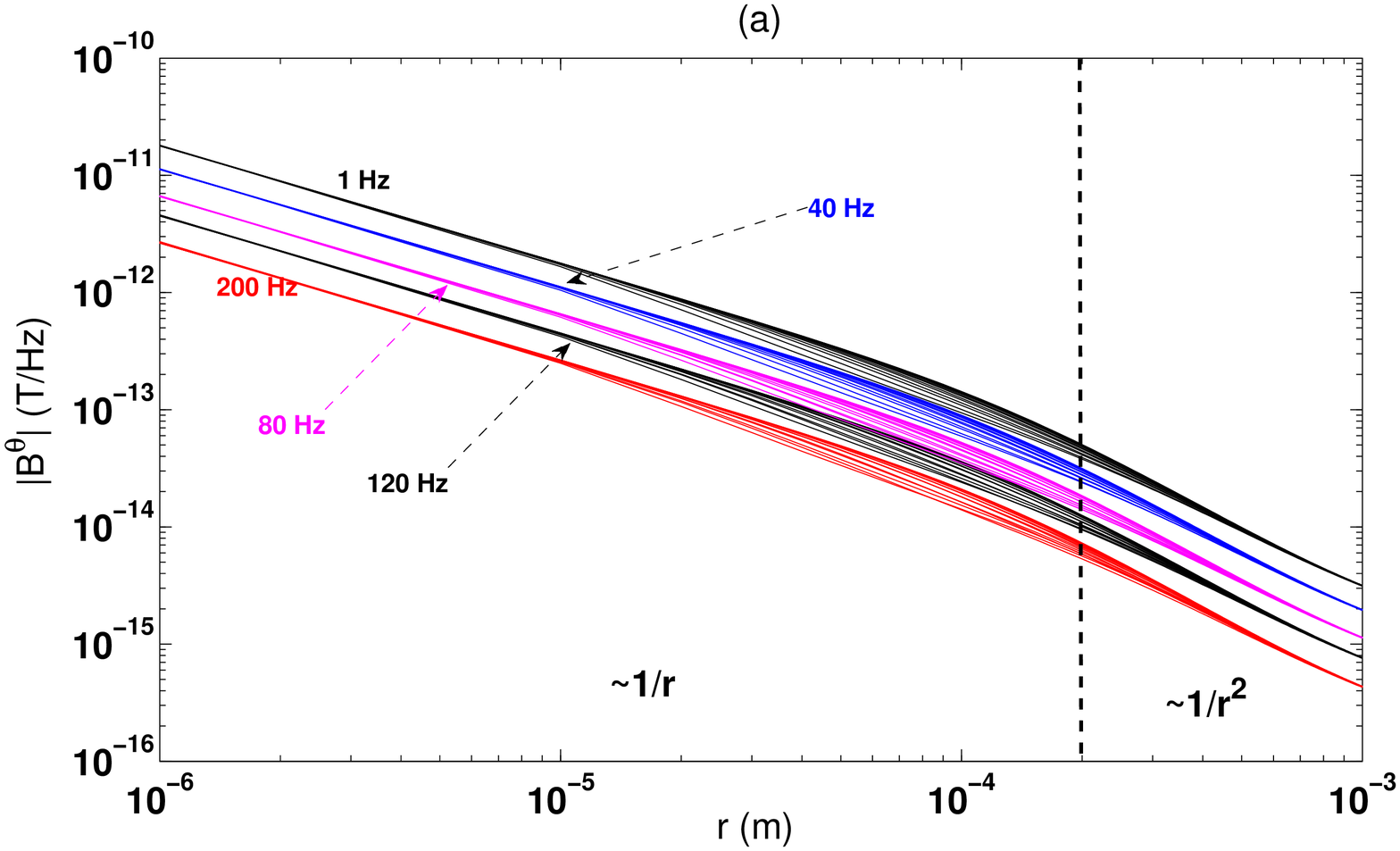}
\includegraphics[width=8cm]{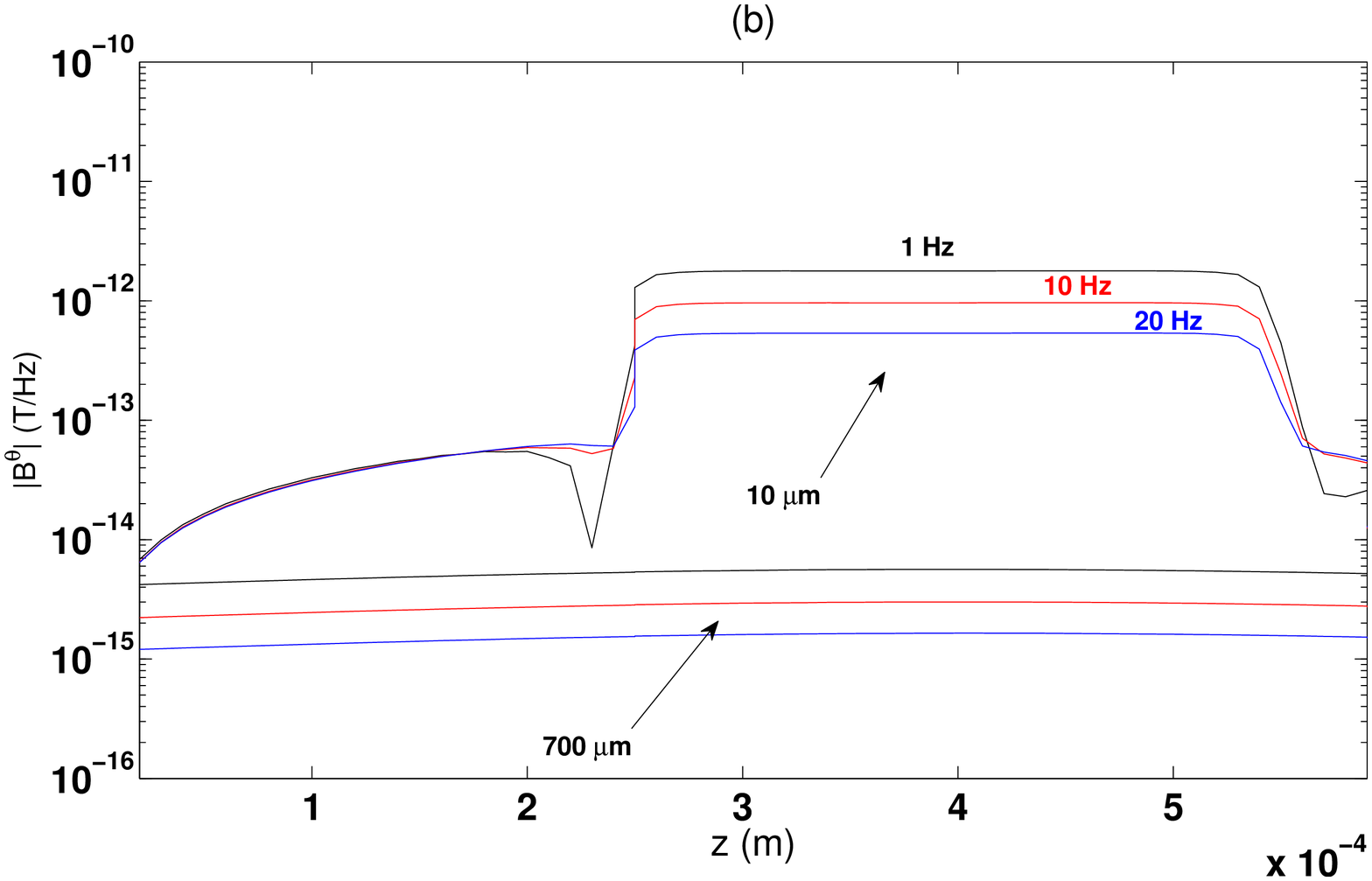}

\caption{(Color online) Distance-dependence of the magnetic induction
  for a ball-and-stick model with resistive media.  The boundary
  conditions are represented in Figs.~\ref{fig10} and \ref{fig11}.
 (a) Attenuation law for
  the modulus of \b{$\textbf{B}^{\theta}$} relative to $r$ (direction
  perpendicular to the axis of the stick).  For \b{$r<100~\mu m =l/6
    $}, the attenuation is varying as $1/r$ with a proportionality
  constant that depends on frequency.  For \b{$r> 200~\mu m =l/3 $},
  the attenuation varies as \b{$1/r^2$} and is roughly independent of
  frequency.  (b) Attenuation law relative to $z$ (direction parallel
  to the axis of the stick).  The attenuation does not depend on
  frequency for positions outside the regions between the synapses.
 In all cases, the phase varied very little and was
  not represented.}

    \label{fig11}
\end{figure}

\subsubsection{Attenuation law with distance in extracellular space}\label{S3.2.3}

\begin{figure}[bht!]
\centering
\includegraphics[width=8cm]{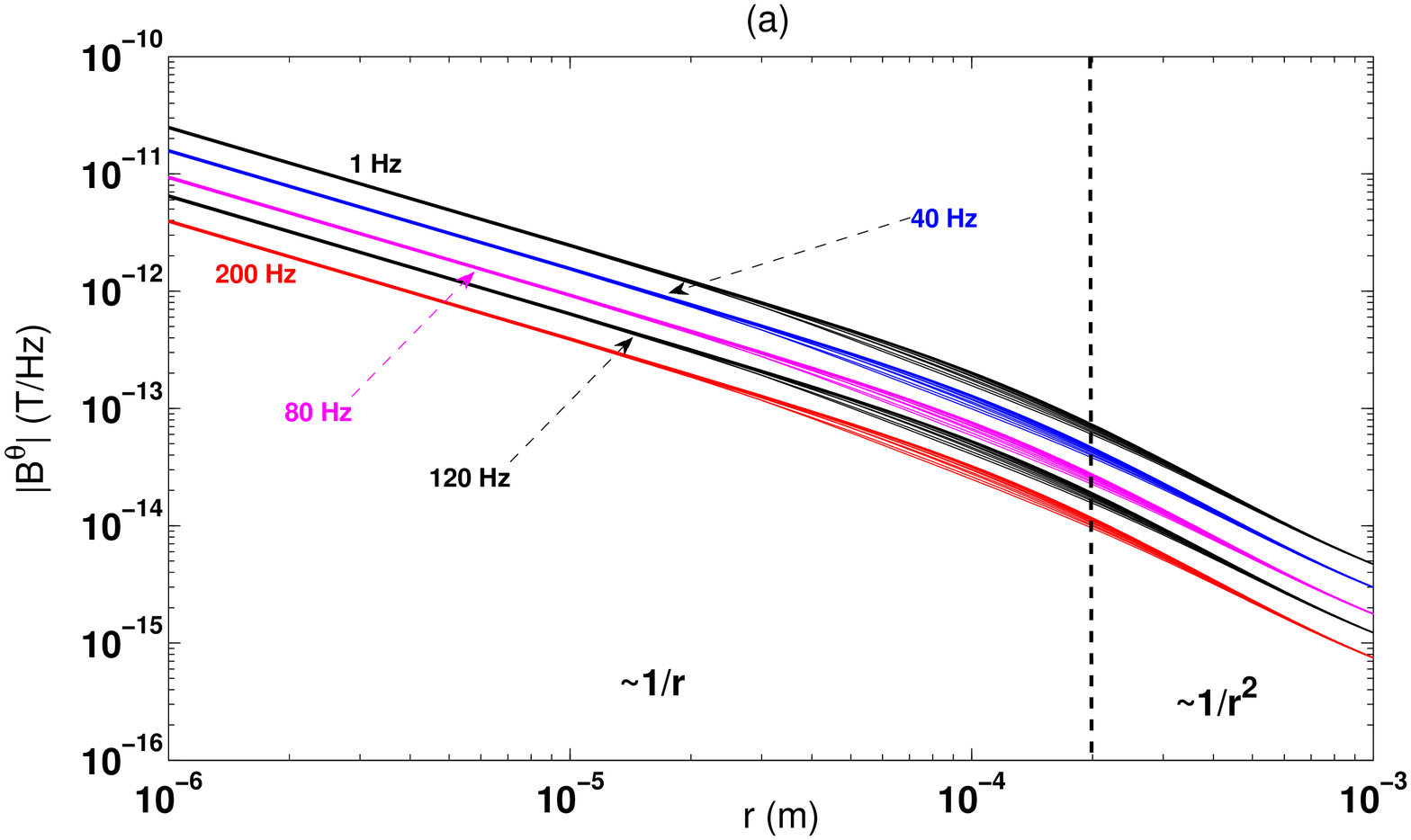}
\includegraphics[width=8cm]{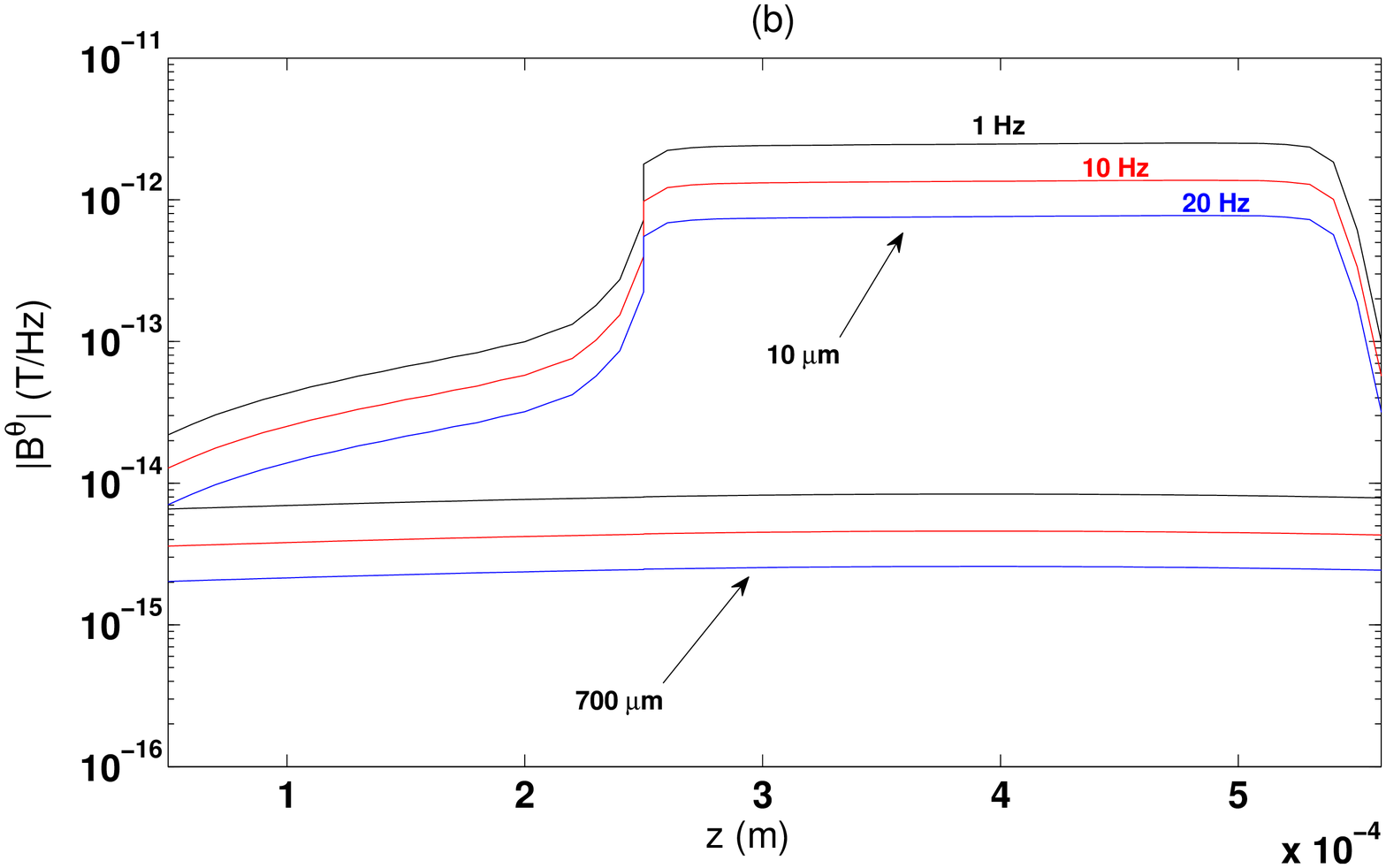}

\caption{(Color online) Distance-dependence of the magnetic induction
  for a ball-and-stick model with diffusive media.  Same arrangement
  as in Fig.~\ref{fig11}, but with boundary conditions as represented
  in Figs.~\ref{fig9} and \ref{fig10}.  (a) Attenuation law for the modulus of
  \b{$\textbf{B}^{\theta}$} relative to $r$.  As for the resistive
  model, the attenuation varies as $1/r$ for \b{$r<100~\mu m =l/6 $},
  and as \b{$1/r^2$} for \b{$r> 200~\mu m =l/3 $}.  (b) Attenuation
  law relative to $z$.  Contrary to the resistive model, the
  attenuation depends on frequency for all positions.  }

    \label{fig12}
\end{figure}

In this section, we show that the attenuation law of \b{$B^{\theta}$}
relative to distance in the extracellular medium (Figs.~\ref{fig11}
and \ref{fig12}) depends on the nature of the extracellular impedance.
Fig.~\ref{fig11} shows an example of the attenuation obtained in a
resistive medium, while Fig.~\ref{fig12} shows the same for a medium
with diffusive properties (Warburg impedance).  The parameters are the
same as for Figs.~\ref{fig7}-\ref{fig8}, and
Figs.~\ref{fig9}-\ref{fig10}, respectively. 

>From Figs.~\ref{fig11} and \ref{fig12}, one can see that the nature of
the extracellular medium has little effect on the attenuation law
relative to distance $r$ for a position $z$ in between the synaptic
sites.  However, the nature of the medium is more influential outside
of this region.  For \b{$r<100~\mu m =l/6 $}, the attenuation varies
as \b{$1/r$} and is dependent on frequency, while for \b{$r> 200~\mu m
  =l/3 $}, the attenuation varies as \b{$1/r^2$}.  The nature of the
medium changes the position dependence of the magnetic induction.  In
a diffusive medium, the ``return current'' more strongly depends on
frequency compared to a resitive medium, and the partial derivative of
\b{$\textbf{B}^{\theta}$} relative to $z$ is less abrupt (low-pass
filter).

When comparing Figures \ref{fig7} to \ref{fig12}, one can see that
the nature of the cytoplasm impedance has a larger effect than the
extracellular impedance.  The intracellular impedance has more effect
on the slope of the frequency dependence of the magnetic induction on
the surface of the neuron (boundary conditions), while the
extracellular impedance affects more the attenuation law with
distance.  The latter effect is due to the fact that the extracellular
impedance affects the return currents, and therefore plays a screening
effect on \b{$\textbf{B}^{\theta}$}, in a frequency-dependent manner.  It is
interesting to see that the nature of the impedances affects
\b{$\textbf{B}^{\theta}$}, although we have roughly the same magnetic
permeability as vacuum.

%-------------------------- DISCUSSION ----------------------

\section*{Discussion}

In this paper, we have derived a cable formalism to calculate the
extracellular magnetic induction \b{$\vec{\textbf{B}}$} generated by
neuronal structures.  A first original contribution of this formalism
is to allow, for the first time, to evaluate \b{$\vec{\textbf{B}}$} in
neurons embedded in media which can have arbitrary complex electrical
properties, such as for example taking into account diffusive or
capacitive effects in the extracellular space.  To this end, it is
necessary to use the ``generalized cable'' formalism indroduced
recently \cite{BedDes2013}, which generalizes the classic Rall cable
formalism \cite{Rall1962,Rall1995} but for neurons embedded in media
with complex electrical properties.  Using this generalized cable, it
was shown that the nature of the medium influences many properties
such as voltage and axial current attenuation \cite{BedDes2013}.
\corr{We show here that it can also influence neuronal magnetic
  fields.}

\corr{To compare with previous approaches, it is important to note
  that} the present formalism is based on a multi-scale mean-field
theory.  We consider the neuron in interaction with the ``mean''
extracellular medium, characterized by a specific impedance
\cite{BedDes2013}.  Using such a formalism, we can study the influence
of the nature of the extracellular medium impedance on the axial
current, and deduce its effect on the spatial and frequency profile of
\b{$\vec{\textbf{B}}$}.  This represents a net advantage over a
classical mean-field theory, where the medium is considered as a
continuum where the biological sources are not explicitly represented.
An alternative approach consists of using the Biot-Savart law in three
dimensions, within a mean-field model of the
cortex~\cite{Hamailainen1993}.  This approach can be considered as a
first-order approximation of the formalism we present here.  However,
it is strictly limited to resistive media, and cannot be used to
investigate the fields generated in non-resistive or non-homogeneous
media, with complex electrical properties.  In such a case, the
present formalism should be used.

\corr{The present formalism can be extended or further developed in
  several ways.  First, some predictions of the formalism can be
  tested experimentally.  Our numerical simulations} show that the
electric nature of intracellular and extracellular media influence
many properties of \b{$\vec{\textbf{B}}$}.  This result may seem
surprising at first sight, because the magnetic field itself is not
filtered by the medium, so we would expect \b{$\vec{\textbf{B}}$} to
be independent of the electrical properties of extracellular space.
However, as mentioned above, these properties influence the membrane
currents and the axial currents in the neuron, and thus, in turn, they
also influence \b{$\vec{\textbf{B}}$}.  So this property constitutes
an important prediction of the present formalism, the nature of the
extracellular medium should affect the frequency dependence of
\b{$\vec{\textbf{B}}$}, which can be measured experimentally.
\corr{For example, according to the present work, the PSD of
  \b{$\vec{\textbf{B}}$} should present a frequency-scaling which
  reflects the frequency-scaling of the impedances of the
  intracellular and extracellular media. This influence may perhaps
  explain the particular frequency scaling observed for MEG
  signals~\cite{Deg2010} (reviewed in \cite{DesBed2013}).  The fact
  that the extracellular impedance influences \b{$\vec{\textbf{B}}$}
  also has clear consequences for the so-called inverse problem of
  finding the neuronal sources from recorded electric or magnetic
  brain signals.  These types of analysis constitute an important
  future development of the present work.}

\corr{A second possible direction for future work is to extend the
  present formalism to simulate complex neuronal morphologies.  We
  have shown here that \b{$\vec{\textbf{B}}$} can be approximated by
  successive analytical iterations.  Such an analytic approach relies
  on the assumption that the continuous cylinder is of constant
  diameter, but it is valid for arbitrarily complex extracellular
  electric properties.  Thus, it should be possible to apply the same
  approach to simulate any complex neuronal morphology, using a set of
  such continuous cylinder compartments.  Because the approximate
  solution is analytic,} this formalism can lead to very efficient
algorithms to simulate the magnetic field generated by complex
neuronal morphologies or populations of neurons.  This also
constitutes a main follow-up of the present work.

It is important to note that some of the previously-proposed models
\corr{of magnetic fields generated by complex neuronal morphologies}
are based on a direct application of the Biot-Savart law
\cite{Murakami,Cassara}, which neglects the return currents and is
equivalent to consider that the neuron is embedded into vacuum.  In
reality, the neuron exchanges currents with extracellular space, and
generates return currents, which also participate to the the genesis
of \b{$\vec{\textbf{B}}$}.  One main advantage of the present
formalism is that these return currents are taken into account, and
thus we believe that it provides a good estimate of the ``net''
magnetic induction \b{$\vec{\textbf{B}}$} generated by neurons
embedded in realistic extracellular media.

\corr{A third possible extension of the present formalism is to
  include the effect of magnetic stimulation.  Because of the recent
  emergence of non-invasive techniques such as the trans-cranial
  magnetic stimulation \cite{George1999}, it is likely that
  understanding the effect of magnetic stimulation will become
  increasingly important in the future.  In our formalism, it is
  possible to integrate this effect from the righthand term of
  Eq.~(\ref{eq9}), because this term takes into account the phenomenon
  of electromagnetic induction.  Evidently, the solution in space will
  be different from what is presented here, because this additional
  term implies \b{$\nabla^2\vec{\textbf{B}} \neq 0$}.  However, most
  of the formalism developed here can be used because we have
  \b{$\vec{E} = -\nabla V -\frac{\partial \vec{A}}{\partial t}$}
  instead of \b{$\vec{E} = -\nabla V $}.  Thus, with minor
  modifications, it is possible to consider the effect of magnetic
  stimulation in neurons, together with the complex properties of the
  extracellular medium, generalizing previous approaches
  \cite{Durand}.  Here again, the effect of magnetic stimulation
  depends on the admittance of the medium, which constitutes another
  way by which neuronal behavior may depend on the electric properties
  of extracellular space.}

%-------------------------- APPENDICES ---------------------
\begin{appendix}

\section*{Appendices}

\section{\corr{Cylindric symmetry and the direction of \b{$\vec{B}$}}}
\label{appA}
\setcounter{equation}{0} 
\numberwithin{equation}{section}
\setcounter{figure}{0} 
\numberwithin{figure}{section}

\corr{In this appendix, we calculate the direction of \b{$\vec{B}$}.
  To do this, we use the expression of the Vector Potential
  \b{$\vec{\textbf{A}}$}
  (\b{$\vec{\textbf{B}}=\nabla\times\vec{\textbf{A}}$}) in conditions
  of Coulomb's Gauge (\b{$\nabla\cdot\vec{\textbf{A}}=0$}) and the law
  of Kelvin-Maxwell
  (\b{$\nabla\cdot\vec{\textbf{B}}=0$}~[Eq.~(\ref{eq1}~iii)]).}

\subsection{\corr{Component \b{$\textbf{B}^z$} on the surface of the
    continuous cylinder compartment }}

\corr{If one substitutes in Eq.~(\ref{eq1}~iv),
\b{$\vec{\textbf{B}}=\nabla\times\vec{\textbf{A}}$} (within Coulomb's Gauge),
we obtain:} \b{
\begin{equation}
\nabla\times\vec{\textbf{B}}=\nabla\times (\nabla\times\vec{\textbf{A}}) \equiv
-\nabla^2\vec{\textbf{A}}+\nabla(\nabla\cdot\vec{\textbf{A}})=
-\nabla^2\vec{\textbf{A}}=\mu_o\vec{\textbf{j}}^{~g} ~ .
\label{eqa1}
\end{equation}} 
\corr{Thus, each component of \b{$\vec{\textbf{A}}$} is solution of a ``Poisson'' 
type equation, and we can write in cylindric coordinates:}
\b{
\begin{equation}
\label{eqa2}
\vec{\textbf{\textbf{A}}}(\vec{x},\omega ) = \frac{\mu_o}{4\pi}\iiint\limits_{\mathcal{E}_t} \frac{\vec{\textbf{j}}^{~g}(\vec{x}~',\omega )}{\sqrt{r^2+r'^2+2rr'cos(\theta-\theta')+(z-z')^2 }}~r'dr'd\theta'dz'
\end{equation}}

\corr{if we assume that \b{$\vec{\textbf{A}}=0$} at infinite distance.  The
integration domain \b{$\mathcal{E}_t$} represents all space.  However,
assuming that the current field in a continous cylinder compartment
follows cylindric symmetry, we can write that in any point of space,
the generalized current density is given by:
\b{$\vec{\textbf{j}}^{~g}= \textbf{j}^{r~g}(r,z)~\hat{e}_r +
  \textbf{j}^{z~g}(r,z)~\hat{e}_z $} where \b{$\hat{e}_z$} and
\b{$\hat{e}_r$} are respectively parallel and perpendicular to the
symmetry axis of the axial current.  It follows that the Vector
Potential is of the form \b{$\vec{\textbf{A}} =
  \textbf{A}^r(r,z)~\hat{e}_r + \textbf{A}^z(r,z)~\hat{e}_z $}
[Eq.~(\ref{eqa2})].  Thus, the component \b{$\textbf{B}^z$} of
\b{$\vec{\textbf{B}}$} is always equal to zero, since we have
\b{$\textbf{B}^z=(\nabla\times\vec{\textbf{A}})^{~z} =0$}.}

\subsection{\corr{The component \b{$\textbf{B}^r$} on the surface of
    the continuous cylinder compartment}}

\corr{The application of Kelvin-Maxwell's law [Eq.~(\ref{eq1}~iii)] implies
that the surface integral (Fig.~\ref{fig2}) of \b{$\textbf{B}^r$} gives:} 
\b{
\begin{equation}
\label{eqa3}
\iint\limits\limits_{\mathcal{S}_B} \textbf{B}^{r} ~dS = 
\iint\limits_{\partial \mathcal{D}} \vec{\textbf{B}}.\hat{n}~ dS \equiv
\iiint\limits_{\mathcal{D}} \nabla\cdot\vec{\textbf{B}}~dv =0
\end{equation}}
\corr{because the component \b{$\textbf{B}^z=0 \Rightarrow
  \iint\limits\limits_{\mathcal{S}_A} \vec{\textbf{B}}\cdot\hat{n}=
  \iint\limits\limits_{\mathcal{S}_C} \vec{\textbf{B}}\cdot\hat{n}
  ~dS=\iint\limits\limits_{\mathcal{S}_A}
  |\textbf{B}^{z}|dS=\iint\limits\limits_{\mathcal{S}_C}
  |\textbf{B}^{z}|dS = 0 $} (for a plane perpendicular to the surface
\b{$\mathcal{S}_B$} of the compartment. Thus, we can deduce that
\b{$\textbf{B}^{r}=0$} because the integral of \b{$\textbf{B}^{r}$} is
zero for a surface of arbitrary length \b{$\mathcal{S}_B$}.
Consequently, the general expression of \b{$\vec{\textbf{B}}$} over
the surface of a continuous cylinder compartment is given by:}
\b{
\begin{equation}
\label{eqa4}
\vec{\textbf{B}}=\textbf{B}^{\theta}\hat{e}_{\theta} ~ .
\end{equation}}
\corr{Note that electromagnetic induction is taken into account in this
derivation because we did not use the explicit value of 
\b{$\nabla\times\vec{\textbf{E}}$} when deriving Eq.~(\ref{eq1}~ii).}

\section{\corr{Solving \b{$\nabla^2\vec{\textbf{B}}=0$} for a continuous
cylinder compartment}}
\label{appB}
\setcounter{equation}{0} 
\numberwithin{equation}{section}
\setcounter{figure}{0} 
\numberwithin{figure}{section}

\begin{figure}[h!] 
\centering
\includegraphics[width=15cm]{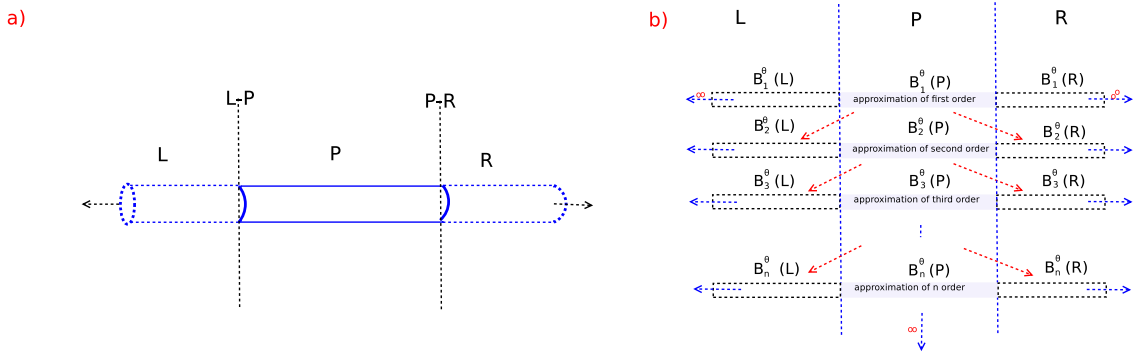}

\caption{\small (Color online) Scheme to calculate
  \b{$\textbf{B}^{\theta}$}.  {\bf a} Extension of the compartment in
  Regions L and R when the principal region (P) consists of a single
  cylinder compartment of radius \b{$a$}.  {\bf b}. Calculation
  scheme.  In a first step, we calculate in Fourier space the field
  \b{$\textbf{B}^{\theta}$} by assuming that the boundary conditions
  on the cylinder are such that:
  \b{$\textbf{B}^{\theta}(a,z<0,\omega)=0$} and
  \b{$\textbf{B}^{\theta}(a,z>l,\omega)=0$} in Regions~L and R. The
  solution of \b{$\nabla^2\vec{\textbf{B}}=0$} obtained in such conditions is called
  the {\it first-order solution} of \b{$\textbf{B}^{\theta}$}.  In a
  second step, we improve the boundary conditions by applying the
  Hankel transform of order 1 relative to \b{$r$}, and the continuity
  principle of the solution at the borders L-P and P-R.  We
  re-evaluate the solution of \b{$\nabla^2\vec{\textbf{B}}=0$} with these new boundary
  conditions to obtain a {\it second-order solution} of
  \b{$\textbf{B}^{\theta}$}.  The same iteration is continued.}

\label{figb1}
\end{figure}
     
\corr{In this appendix, we introduce an iterative method to calculate the
solution of \b{$\nabla^2\vec{\textbf{B}}=0$} for a continuous cylinder
compartment of radius \b{$a$} and length \b{$l$}, when the values of
\b{$\textbf{B}^{\theta}$} at its surface are known. We approach the
solution analytically by using the mathematical relations between the
different currents present in the neuron, and the magnetic induction
that these currents produce.}

\corr{We calculate \b{$\textbf{B}^{\theta}$} in space assuming that Region~P
contains only one continuous cylinder compartment [Fig~\ref{figb1}~a],
but the method can be easily generalized to the case with several
compartments [Fig~\ref{figb1}~b].  In this case, to generalize to
\b{$N_p$} compartments, one must determine the boundary conditions of
\b{$\textbf{B}^{\theta}$} for each compartment inside Region~P.  In
all cases, we assume that \b{$\textbf{B}^{\theta}$} satisfies: 1)
\b{$\textbf{B}^{\theta}$} is a continuous function on the borders L-P
and P-R [Fig.~\ref{figb1}]; 2) \b{$\textbf{B}^{\theta}=0$} at infinite
distance; 3) \b{$\textbf{B}^{\theta}=0$} on the symmetry axis of the
compartment.}

\corr{To calculate the solution, we extend the original compartment in
Regions L and R using the same radius \b{$a$} (Fig.~\ref{figb1}a).  Note
that by convention, we place the symmetry axis on the z-axis, and
place the continuous cylinder between coordinates \b{$z=0$} and
\b{$z=l>0$}.}

\corr{At the first order of the iteration, we assume that}
\b{
$$
\left \{
\begin{array}{ccccc}
\textbf{B}^{\theta}(a,z<0,\omega) 
&=& 0 \\\\
\textbf{B}^{\theta}(a,z>l,\omega)
&=& 0
\end{array}
\right .
$$}
\corr{over the surfaces of the extended compartment (L and R), which ensures
the spatial continuity of the first-order solution at the borders L-P
and P-R. \textit{ A priori} this choice is arbitrary but we have
chosen here a particular attenuation law which neglects the radius of
the extended compartment.  Following this first choice, we calculate
the solution of Eq.~(\b{$\ref{eq9}$})  by using complex Fourier
transform along $z$.  This leads to:} 
\b{
\begin{equation}
\textbf{B}^{\theta}(r,z,\omega)= \frac{1}{2\pi}
\int_{-\infty}^{+\infty}g(r,k_z,\omega )~e^{+ik_zz}~dk_Z ~ ,
\label{eqb1}
\end{equation}}
\corr{where}
\b{  
\begin{equation}
g(r,k_z,\omega )= 
\int_{-\infty}^{+\infty}\textbf{B}^{\theta}(r,z,\omega)~e^{-ik_zz}~dz ~ .
\label{eqb2}
\end{equation}}
\corr{We next substitute Eq.~(\ref{eqb1}) in Eq.~(\ref{eq9}), which
leads to}
\b{
\begin{equation}
\int_{-\infty}^{+\infty}
[\frac{d^2g}{dr^2}+\frac{1}{r}\frac{dg}{dr}-(k_z^2+\frac{1}{r^2})~g]~e^{+ik_zz}dz =0 ~ .
\label{eqb3}
\end{equation}}
\corr{when electromagnetic induction is neglected.}

\begin{figure}[h!] 
\centering
\includegraphics[width=12cm]{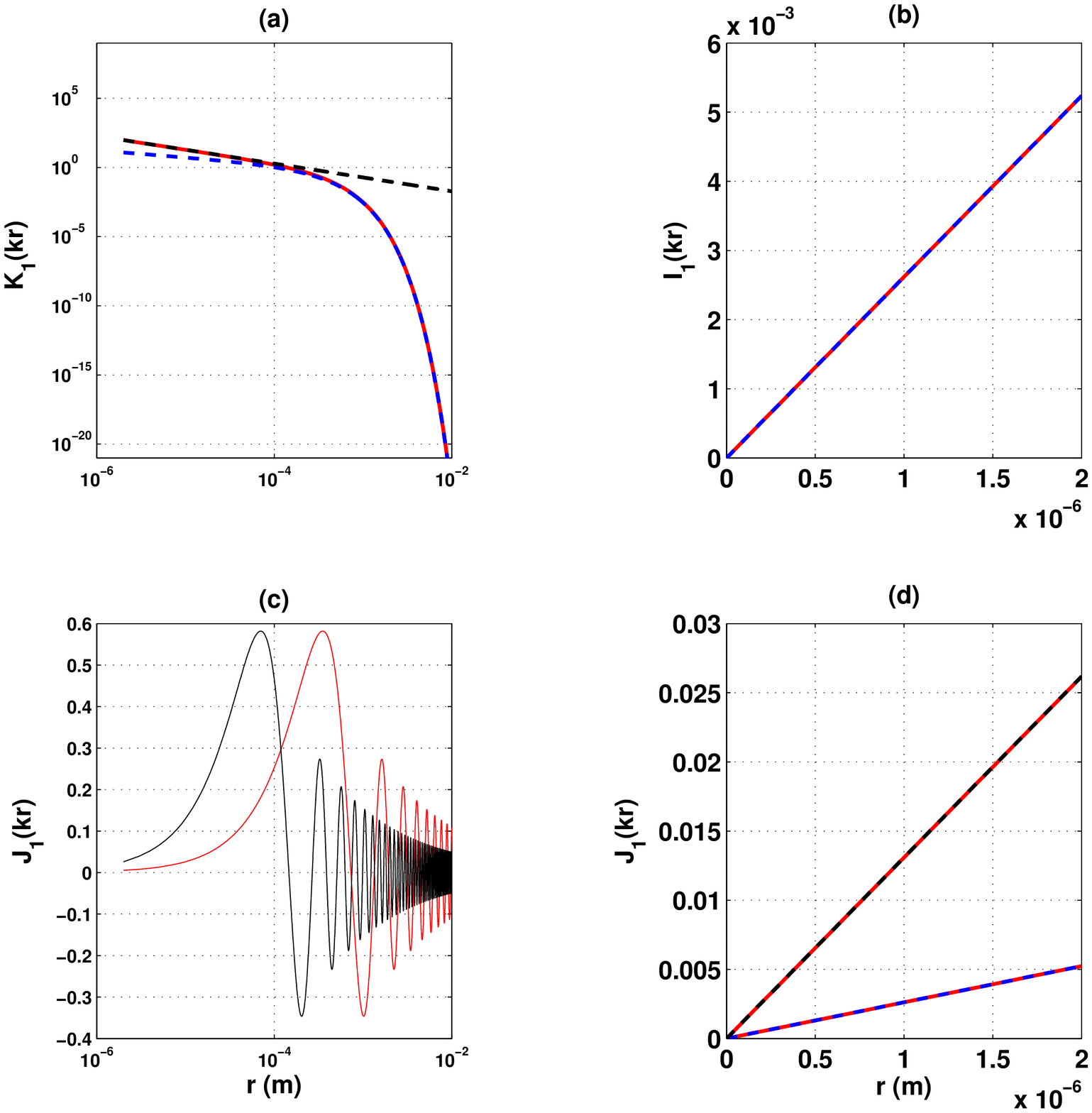}

\caption{\small (Color online) Bessel functions for a continuous
  cylinder compartment.  The Bessel functions are indicated as a
  function of the distance \b{$r$} perpendicular to the axis of the
  cylinder; the cylinder had a length \b{$l$} of \b{$300~\mu m$} and a
  radius \b{$a=2~\mu m$}. We have \b{$k'=2\times 10^4~m^{-1}$}. {\bf
    a}. \b{$K_1$} as a function of distance.  The red curve shows the
  function \b{$K_1(k r)$} with \b{$k=k'$}. The black dashed straight
  line represents the function \b{$ 1/kr$} and the blue dashed curve
  represents the asymptotic behavior of \b{$K_1$} for \b{$r
    \rightarrow \infty$}. We have
  \b{$K_1(kr)\overset{\infty}{\longrightarrow}\sqrt{\frac{\pi}{2kr}}~e^{-kr}$}
  .  At short distances (smaller than \b{$2/k$}), the function
  \b{$K_1$} decays linearly with distance, but for large distances
  (\b{$r>\pi/k$}), it converges more rapidly than an exponential decay
  with distance.  {\bf b}. The function \b{$I_1$} (modified Bessel
  function of first kind of order 1) is well approximated by a
  straight line (\b{$I_1(k r)=k r/2$}) for \b{$k=k'$} when \b{$r<a$}.
  {\bf c}. Bessel function of first kind of order 1 when \b{$r >a$}.
  The blue and black curves correspond respectively to \b{$k =k'$} and
  \b{$k =5k'$}.  Note that we have \b{$J_1(k r)
    \overset{\infty}{\longrightarrow} \sqrt{\frac{2}{\pi
        kr}}~cos(kr-\frac{3\pi}{4})$}. We see that when \b{$k$} is
  large enough, the function \b{$J_1(kr)$} can capture small spatial
  variations. {\bf d.} The Bessel function of first kind of order 1 is
  equivalent to a straight line (\b{$J_1(kr)= kr/2 $}) when \b{$r<a$}.
  The blue dashed curve corresponds to approximating \b{$J_1(kr)$} by
  a linear law for \b{$k = k'$}, while the black dashed curve is the
  linear approximation for \b{$k =5k'$}.  The red curves correspond to
  \b{$J_1(kr)$} for \b{$k = k'$} and \b{$k = 5k'$}
  \cite{Bateman_1,Bateman_2}.}

\label{figb2}
\end{figure}

\begin{figure}[h!] 
\centering
\includegraphics[width=15cm]{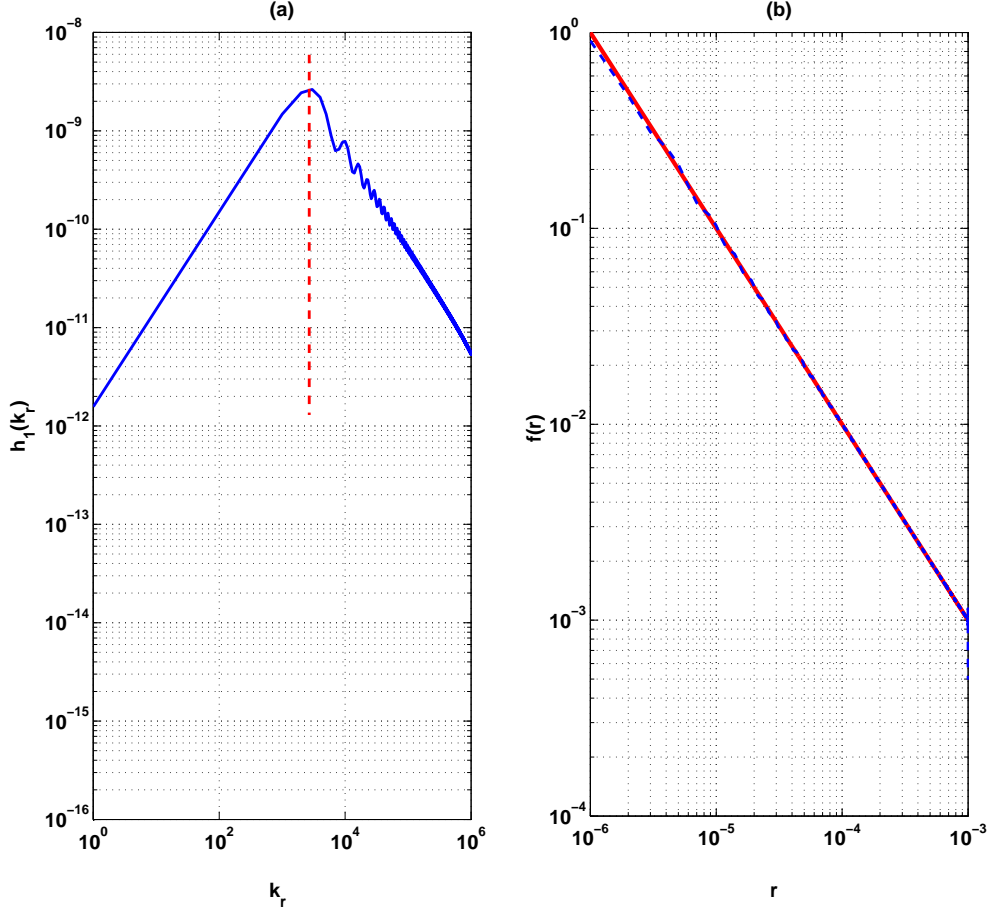}

\caption{\small (Color online) Example of application of the Hankel
  transform of order 1.  {\bf (a).} Approximation using the Hankel
  transform of order 1 of the function \b{$f(r)=H(r-a)1/r$} with \b{$a
    = 1~\mu m$}, \b{$1<k_r< 5\times 10^6 $} and \b{$\Delta k_r
    =10^3$}. The values smaller than \b{$10^3$} are not significant
  because of the value of \b{$\Delta k_r$} is larger than \b{1000}.  We
  can see that the approximation using the Hankel transform is valid
  for distances up to \b{$1~mm$}. {\bf (b).}  Inverse transform
  applied to this approximation (in blue), and comparison with the
  original function (in red), between \b{$1~\mu m $} and \b{$1~mm $}.
  The parameter \b{$k_r$} of the Hankel transform plays a similar role
  as the wave number (\b{$\frac{2\pi}{\lambda}$}) in spatial Fourier
  transform.  The larger \b{$k_r$}, the more sensitive to fine spatial
  details.}
 
\label{figb3}
\end{figure}

\corr{Thus, we have (for \b{$k_z$} and \b{$\omega$} fixed) the following
equality:}
\b{
\begin{equation}
\frac{d^2g}{dr^2}+\frac{1}{r}\frac{dg}{dr}-(k_z^2+\frac{1}{r^2})~g=0
\label{eqb4}
\end{equation}}
\corr{because the Fourier transform of zero is zero. It follows that the
function \b{$g$} must be solution of the modified Bessel differential 
equation of order 1.  The general solution of such an equation is 
given by:}
\b{
\begin{equation}
g(r,k_z,\omega) = 
  c(k_z,\omega)~ I_1~(|k_z|r) +
 d(k_z,\omega)~ K_1~(|k_z|r) ~ ,
 \label{eqb5}
\end{equation}}
\corr{where \b{$I_1$} is a a modified Bessel function of first kind of order
  1 and \b{$K_1$} is a modified Bessel function of second kind of order
  1\footnote{We have \b{$J_1(ir') = iI_1(r')$} and \b{$Y_1(ir') =
      I_1(r') + \frac{2}{\pi}iK_1(r')$}, where \b{$J_1$} is the modified
    Bessel function of first kind of order 1, and \b{$Y_1$} is the
    Bessel function of second kind of order 1.}.  Such functions are
  illustrated in Fig.~(\ref{figb2}) as a function of $r$ for typical
  parameter values that corresponds to neurons.  Note that we must
  assume that \b{$k_z$} is very small for function \b{$K_1(|k_z|r)$} to
  have a significant value for large \b{$r$}.  For a fixed value of
  \b{$K_1$}, we have \b{$k_z \sim 1/r$}.  Note that, for the 
  typical geometrical size of neurons and distances studied here 
  ($<1~mm $), the valuesof $k_r$ are pertinent when they are larger 
  than 1000~$m^{-1}$ [Fig. \ref{figb3}].}

\corr{Finally, to evaluate the coefficients \b{$c(k_z,\omega)$} and
\b{$d(k_z,\omega)$}, we apply the continuity condition of
\b{$\textbf{B}^{\theta}$} between the interior and exterior of the extended
compartment, and that \b{$\textbf{B}^{\theta}$} must be zero on the symmetry
axis of the compartment (\b{$r=0$}), as well as at infinite distance.
Because \b{$|I_1~(\infty,\omega)|=\infty$}, we must assume that
\b{$c(k_z,\omega)=0$} outside of the cylinder, and because
\b{$|K_1~(0,\omega)|=\infty$}, we must assume that
\b{$d(k_z,\omega)=0$} inside of the cylinder.  Taking these conditions
into account, we obtain:} \b{
\begin{equation}
\left \{
\begin{array}{cccccc}
exterior & r \geq a & d(k_z,\omega) & = & \frac{g(a,k_z,\omega)}{K_1(|k_z|a)} 
\\\\
interior & r\leq a & c(k_z,\omega) & = & \frac{g(a,k_z,\omega)}{I_1(|k_z|a)} 
\end{array}
\right .
\label{eqb6}
\end{equation}}
\corr{It follows that the approximative solution of first-order is given by:}
\b{
\begin{equation}
\left \{
\begin{array}{cccccc}
exterior & r \geq a & \textbf{B}^{\theta}(r,z,\omega) & = & \frac{1}{2\pi}\int_{-\infty}^{+\infty}g(a,k_z,\omega)\frac{K_1(|k_z|r)}{K_1(|k_z|a)}~e^{+ik_zz}~dk_z 
\\\\
interior & r \leq a & \textbf{B}^{\theta}(r,z,\omega) & = & \frac{1}{2\pi}\int_{-\infty}^{+\infty} g(a,k_z,\omega)\frac{I_1(|k_z|r)}{I_1(|k_z|a)}~e^{+ik_zz}~dk_z  
\end{array}
\right . ~ ,
\label{eqb7}
\end{equation}}
\corr{where the function \b{$g(a,k_z,\omega)$} is given by Eq.~(\ref{eqb2}):}
\b{
\begin{equation}
g(a,k_z,\omega)= \int_{-\infty}^{+\infty}\textbf{B}^{\theta}(a,z,\omega)
~e^{-ik_zz}~dz~.
\label{eqb8}
\end{equation}} 

\corr{This first iteration gives us a first-order approximation of
\b{$\textbf{B}^{\theta}$}, which is refined in successive iterations,
as schematized in Fig.~\ref{figb1}~(b).  We use the first-order
approximation in Region~P to calculate the solutions in Regions L and
R.  To do this, we use the first-order Hankel transform\footnote{This
  is equivalent to the first-order Fourier-Bessel transform. This
  particular transform was chosen here because the function
  \b{$J_1(k_r r)$} has the same boundary conditions as in the present
  problem [Fig.~\ref{figb2}~(c-d)]: it is equal to zero for
  \b{$r=0$} and for \b{$r\rightarrow\infty$}.}  for the variable
\b{$r$}.  To do this, one applies the continuity principle at the
borders L-P and P-R. This gives the following relations:} \b{
\begin{equation}
\textbf{B}^{\theta}(r,z,\omega) = \int_0^{\infty} k_r h_1(k_r,z,\omega)J_1(k_rr)~dk_r
\label{eqb9}
\end{equation}}
\corr{where}
\b{
\begin{equation}
h_1(k_r,z,\omega) = \int_0^{\infty} r \textbf{B}^{\theta}(r,z,\omega)~J_1(k_rr)~dr
\label{eqb10}
\end{equation}}
\corr{in Regions \b{$L$} and \b{$R$}.}

\corr{The Hankel tranform is a calculus technique similar to the wavelet
transform [Fig.~\ref{figb2}~(c-d)]. Note that the values of \b{$k_r$} vary
inversely proportional to the values of \b{$r$}, similarly to the
relation between parameter \b{$k_z$} and \b{$z$} above [Fig.~\ref{figb3}].}

\corr{By substituting Eq.~(\ref{eqb9}) into Eq.~(\ref{eq9}), and neglecting
electromagnetic induction as above, we obtain for fixed \b{$\omega$}:}
\b{
\begin{equation}
 \int_0^{\infty} \{~[r^2 \frac{d^2 J_1(k_rr)}{dr^2}
+r \frac{d J_1(k_rr)}{dr}-J_1(k_rr)]~h_1(k_r,z,\omega) +r^2\frac{d^2h_1(k_r,z,\omega)}{d z^2}
~\}~k_r~dk_r =0
\label{eqb11}
\end{equation}}
\corr{In addition, the first-order Bessel function satisfies the following
equation:} \b{
\begin{equation}
r^2 \frac{d^2J_1(k_rr)}{dr^2} 
+ r\frac{dJ_1(k_rr)}{dr} +  [k_r^2r^2 -1]J_1(k_rr)
= 0 ~ .
\label{eqb12}
\end{equation}}
\corr{It follows that}
\b{
\begin{equation}
 \int_0^{\infty} r^2[ \frac{d^2 h_1(k_r,z,\omega)}{d z^2}-k_r^2h_1(k_r,z,\omega)]J_1(k_rr)~k_r~dk_r =0 ~ .
 \label{eqb13}
\end{equation}}
\corr{Because the Hankel transform of zero is zero, we can write for fixed 
values of \b{$k_r$} and \b{$\omega$}:}
\b{
\begin{equation}
\frac{d^2 h_1}{d z^2}
-k_r^2~h_1=0 ~ .
\label{eqb14}
\end{equation}}
\corr{Thus, the general solution of Eq.~(\ref{eqb14}) is given by:} \b{
\begin{equation}
h_1(k_r,z,\omega) = a(k_r,\omega) e^{+k_rz} + b(k_r,\omega)e^{-k_rz} +c(k_r,\omega)k_rz + d(k_r,\omega) ~ .
\label{eqb15}
\end{equation}}
\corr{Using the condition that \b{$\textbf{B}^{\theta}$} vanishes at infinite 
distance for each frequency, implies that, for each frequency, \b{$a=c=d=0$} 
when \b{$z>l$}, and \b{$a=b=d=0$} when \b{$z<0$}. Consequently, the solution 
in Regions \b{$L$} and \b{$R$} are given by:}
\b{
\begin{equation}
\left \{
\begin{array}{cccccc}
\textbf{B}^{\theta}(r,z,\omega) &=& \int_0^{\infty} h_1^L(k_r,\omega)J_1(k_rr)~ke^{-k_r|z|}~dk_r &  z<0
\\\\
\textbf{B}^{\theta}(r,z,\omega) &=& \int_0^{\infty} 
h_1^R(k_r,\omega)J_1(k_rr)~ke^{-k_r|z-l|}~dk_r & z >l
\end{array}
\right . ~ ,
\label{eqb16}
\end{equation}}
\corr{where \b{$h_1^i$} for \b{$i=L$} and \b{$i=R$} are given by the continuity 
conditions at \b{$z=0$} and \b{$z=l$}, and we obtain: }
\b{
\begin{equation}
\left \{
\begin{array}{cccccc}
h_1^L(k_r,\omega)&=& h_1(k_r,0,\omega)  &=& \int_0^{\infty} r\textbf{B}^{\theta}(r,0,\omega)J_1(k_rr)~dr \\\\
h_1^R(k_r,\omega)&=&h_1
(k_r,l,\omega) &=& \int_0^{\infty} r\textbf{B}^{\theta}(r,l,\omega)J_1(k_rr)~dr
\end{array}
\right .
\label{eqb17}
\end{equation}}
\corr{It follows that we can calculate the new limit conditions on the extended
compartment, by applying Eqs.~(\ref{eqb1}). We obtain: }
\b{
\begin{equation}
\left \{
\begin{array}{cccccc}
\textbf{B}^{\theta}(a,z,\omega) &=& \int_0^{\infty} h_1^L(k_r,\omega)J_1(k_ra)~ke^{-k_r|z|}~dk_r &  z<0
\\\\
\textbf{B}^{\theta}(a,z,\omega) &=& \int_0^{\infty} 
h_1^R(k_r,\omega)J_1(k_ra)~ke^{-k_r|z-l|}~dk_r & z >l
\end{array}
\right . ~ ,
\label{eqb18}
\end{equation}}

\corr{After applying the Hankel transform of first-order, if we recover the
same boundary conditions that were assumed at the borders of the
cylinder compartment, then we have reached the exact solution.  If
this is not the case, we can continue to improve the approximation of
the solution by further iterations [Fig.~\ref{figb1}].  To do this, one
considers the original boundary conditions in Region~P together with
the new expressions for the boundary conditions at the extended
compartment (L and R) according to Eqs.~(\ref{eqb18}). One applies the
complex Fourier transform on axis $z$ [Eqs. (\ref{eqb7}) and
(\ref{eqb8})] to obtain a higher-order approximation.  The iteration
is then continued until one obtains a satisfactory solution
[Fig.~\ref{figc1}.1].}

\section{Convergence of the iterative method 
  \label{appC}}

\setcounter{equation}{0} 
\numberwithin{equation}{section}
\setcounter{figure}{0} 
\numberwithin{figure}{section}

In this appendix, we show that the iterative method of
Appendix~(\ref{appB}) converges to a unique solution.  We show that
the series of successive approximations of \b{$\textbf{B}^{\theta}$}
increase monotonically and are bounded, which is sufficient to prove
convergence.

At every cycle of the iteration, the Laplace equation is solved, which
gives a approximation of for \b{$\textbf{B}^{\theta}$}.  By virtue of the
theorem of extremum solutions of the Laplace equation
\cite{Smirnov2,Smirnov5}, we can say that the minimum and maximum
values of the real and imaginary parts of the Fourier transform (in
time) of \b{$\textbf{B}^{\theta}$} are necessarily on the surface of
the continuous cylinder compartment (or its extension), for a
transform along the $z$ axis.  Similarly, for a transform along the
$r$ axis, they are necessarily on that surface or at infinite.  It
follows that if \b{$\textbf{B}^{\theta}=f+ig$} on the surface of the
cylinder (or its extension) or at the L-P and P-R interfaces, then we
have \b{$|f_1|\geq |f_2|$} and \b{$|g_1|\geq |g_2|$} at every point in
space when these inequalities are satisfied over the boundary
conditions.
Therefore, the absolute value of real and imaginary parts of the solution, as well as
its modulus, of the first-order solution
\b{$\textbf{B}_1^{\theta}=f_1+ig_1$} are larger or equal to that of
the  solution \b{$\textbf{B}_2^{\theta}=f_2+ig_2$}.  If
this was not the case in a given point $p$, it would be in
contradiction with the extremum value theorem, because Laplace
equation is linear.  Indeed, the difference between the solutions
\b{$\textbf{B}_2^{\theta}-\textbf{B}_1^{\theta}$} is also solution of
Laplace equation for the boundary conditions
\b{$(f_1-f_2)+i(g_1-g_2)$}.  Consequently, the real and imaginary
parts of the solution cannot become negative if the boundary
conditions are positive.

\begin{figure}[bht!]
\label{figc1}
\centering
  \includegraphics[width=10cm]{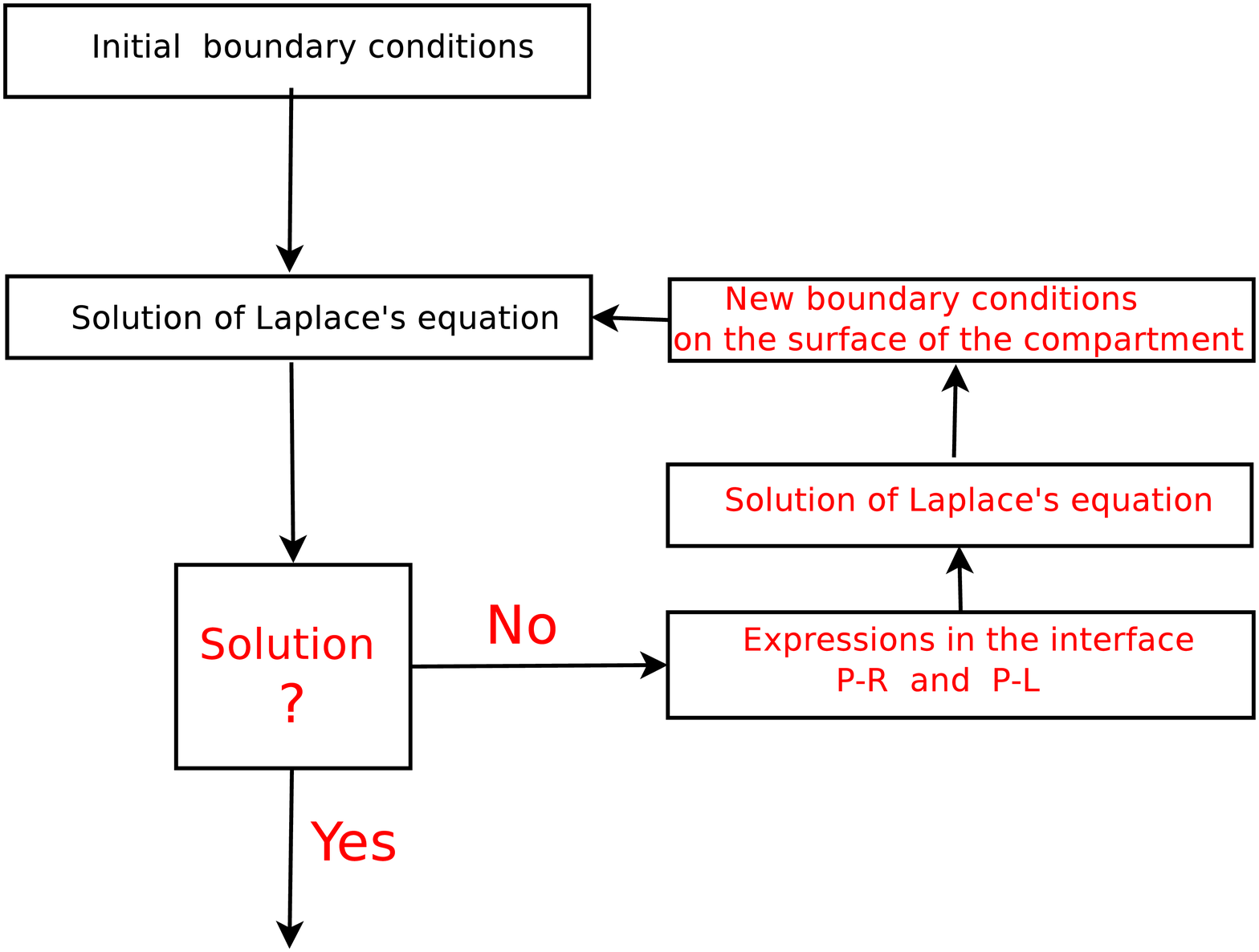}  

  \caption{\small (Color online) Organigram of the iterative method to
    calculate \b{$\textbf{B}^{\theta}$}.}

\end{figure}

To demonstrate that the absolute real and imaginary values are growing
at each iteration [Fig~\ref{figb1} and \ref{figc1}.1], we first
calculate the solution using the Fourier transform along $z$, but
assuming that, on the surface of the extended compartment,
\b{$\textbf{B}^{\theta}$} is zero.  In a second step, we calculate the
solution using the Fourier transform along $r$ and the continuity
principle at the borders L-P and P-R. This second calculation gives
new boundary conditions on the extended cylindric compartment.  These
boundary conditions have real and imaginary values which are necessary
larger or equal (in absolute value) than the ones given for zero
boundary conditions, because the finite length of Region~P is now
taken into account on the surface of the extended compartment.  Thus,
according to above, the modulus of the second-order solution
(calculated using the Fourier transform along $z$) is necessarily
larger than that of the first-order solution, at every point in space.
This reasoning will also apply to the second-order solution because
the extremum value theorem implies that the modulus of the
second-order approximation is larger than the modulus of the
first-order approximation at every point of the interfaces L-P and P-R
[Fig.~\ref{figb1}].  It follows that applying the Fourier transform
with respect to $r$ gives larger values of the boundary conditions for
every point compared to the preceding order, and so on...
Consequently, these successive approximations produce a series of
monotonically increasing values of the modulus of
\b{$\textbf{B}^{\theta}$} at every point of space.  This remarkable
property is a consequences of the theorem of extremum solutions of
Laplace equation.

Finally, we show that this series is bounded.  Indeed, the first-order
solution has real and imaginary values smaller than the solution with
a finite compartment, because \b{$\textbf{B}^{\theta}= 0$} on the
extended compartment.  Thus,
according to the extremum value theorem of Laplace equation, we can
write that for every point in space, the modulus of the first-order
solution is smaller or equal to the exact solution of a single
compartment with no extension. It follows that, for every point in
space, the modulus of the first-order solution of
\b{$\textbf{B}^{\theta}$} is bounded by the modulus of the exact
solution of the compartment with no extension.  This is also valid for
the second-order solution, and so on~...  Consequently, the method
converges to a unique solution in every point in space because we have
a series which is growing and which is bounded.  The unicity of
Laplace equation solution insures that the series converges towards
the exact solution of the compartment without extension.

\end{appendix}

%---------------------------------------------------------------------

%-------------------------- ACKNOWLEDGMENTS ---------------------

\subsection*{Acknowledgments}

Research supported by the CNRS, and grants from the ANR (Complex-V1)
and the European Union (BrainScales FP7-269921, Magnetrodes FP7-600730
and the Human Brain Project).

%-------------------------- REFERENCES ---------------------
%\clearpage
\label{Bibliographie-Debut}

\end{document}